\documentclass[12pt]{article}
\usepackage{epsfig,amsmath}
\usepackage{xspace}
\usepackage{color}
\usepackage{times}
\usepackage{fullpage}
\usepackage{setspace}
\usepackage{epstopdf}
\usepackage{graphicx}
\usepackage{graphics}
\usepackage{subfigure}
\usepackage{bm}
\usepackage{caption}
\usepackage{url}

\newcommand{\superscript}[1]{\ensuremath{^\textrm{\tiny{#1}}}}

\newcommand{\commentout}[1]{}

\long\def\symbolfootnote[#1]#2{\begingroup%
\def\thefootnote{\fnsymbol{footnote}}\footnote[#1]{#2}\endgroup}
\def\blfootnote{\xdef\@thefnmark{}\@footnotetext}

\title{DeepPicker: a Deep Learning Approach for Fully Automated Particle Picking in Cryo-EM}

\author{
 \renewcommand{\thefootnote}{\tiny{\arabic{footnote}}}
 Feng Wang\footnotemark[3]\ \superscript{,}$^{\footnotesize{\dagger}}$
 \renewcommand{\thefootnote}{\tiny{\arabic{footnote}}},
 Huichao Gong\superscript{2}\superscript{,}$^{\footnotesize{\dagger}}$
 \renewcommand{\thefootnote}{\tiny{\arabic{footnote}}},
 Gaochao Liu\superscript{1}\superscript{,}\superscript{4}\superscript{,}$^{\footnotesize{\dagger}}$,\\
 \renewcommand{\thefootnote}{\tiny{\arabic{footnote}}}
 Meijing Li\superscript{1}\superscript{,}\superscript{4}
 \renewcommand{\thefootnote}{\tiny{\arabic{footnote}}},
 Chuangye Yan\superscript{1}\superscript{,}\superscript{5}
 \renewcommand{\thefootnote}{\tiny{\arabic{footnote}}},
 Tian Xia\footnotemark[3]\ \superscript{,}$^{\footnotesize{*}}$,
 Xueming Li\footnotemark[1]\ \superscript{,}\superscript{4}\superscript{,}\superscript{5}\superscript{,}$^{\footnotesize{*}}$,\\
 Jianyang Zeng\superscript{2}\superscript{,}$^{\footnotesize{*}}$\\
}

\begin{document}
\maketitle

\footnotetext[1]{School of Life Sciences, Tsinghua University, Beijing, 100084, China}
\footnotetext[2]{Institute for Interdisciplinary Information Sciences, Tsinghua University, Beijing, 100084, China}
\footnotetext[3]{School of Electronic Information and Communications, Huazhong University of Science and Technology, Wuhan 430074, China}
\footnotetext[4]{Beijing Advanced Innovation Center for Structure Biology, Tsinghua University, Beijing, 100084, China}
\footnotetext[5]{Tsinghua-Peking Joint Center for Life Sciences, Tsinghua University, Beijing, 100084, China}

 \vspace{-11pt}

\symbolfootnote[0]{\hspace{-0.1in} \footnotesize{$\dagger$}These authors contributed equally to this work.}

\symbolfootnote[0]{\hspace{-0.1in} $^*$Corresponding author: Jianyang Zeng,
zengjy321@tsinghua.edu.cn; Xueming Li, lixueming@mail.tsinghua.edu.\\cn; Tian Xia, isutian@gmail.com.}

\vspace{-31pt}
\begin{abstract}
Particle picking is a time-consuming step in single-particle analysis and often requires significant interventions from users, which has become a bottleneck for future automated electron cryo-microscopy (cryo-EM). Here we report a deep learning framework, called DeepPicker, to address this problem and fill the current gaps toward a fully automated cryo-EM pipeline. DeepPicker employs a novel cross-molecule training strategy to capture common features of particles from previously-analyzed micrographs, and thus does not require any human intervention during particle picking. Tests on the recently-published cryo-EM data of three complexes have demonstrated that our deep learning based scheme can successfully accomplish the human-level particle picking process and identify a sufficient number of particles that are comparable to those manually by human experts. These results indicate that DeepPicker can provide a practically useful tool to significantly reduce the time and manual effort spent in single-particle analysis and thus greatly facilitate high-resolution cryo-EM structure determination.
\end{abstract}

\vspace{-10pt}
\section{Introduction}

Recent advance of the single-particle electron cryo-microscopy (cryo-EM) technique has been revolutionizing the structural biology field~\cite{ReviewSH, ReviewCheng, PrimerCheng}, and enabled protein complex structure determination at near-atomic resolution~\cite{TRPV1, Spliceosome, VATPase, DHPR}. However, the high-resolution cryo-EM studies of molecular complexes generally require the selection of a tremendous number of (e.g., hundreds of thousands of) high-quality particles from micrographs. This particle picking step is a labor-intensive step in single-particle data analysis and is a major obstacle for automated cryo-EM pipeline. In the past, particles from cryo-EM micrographs are often selected manually. Such a manual picking process is usually a laborious, tedious and time-consuming task which inevitably requires a considerable amount of human effort to obtain a sufficient number of good-quality particles to ensure high-resolution 3D reconstruction. In addition, manual particle selection is normally subjective and can easily introduce bias and inconsistency due to change in human judgement over time.

To relieve the bottleneck in single-particle data analysis, numerous computational approaches have been proposed to facilitate the particle picking process~\cite{Review2001, Review2004, semisupervised,Chen2007, Hall2004, Huang2004,Adiga2004,DoG,Sorzano2009,PPSVM,TMaCS, MAPPOS}. These methods can be basically divided into three categories, including generative~\cite{Chen2007, Hall2004, Huang2004}, unsupervised~\cite{Adiga2004, DoG} and discriminative approaches~\cite{semisupervised, Sorzano2009, PPSVM}. The generative approaches~\cite{Chen2007, Hall2004, Huang2004} usually measure the similarity to a reference to identify particle candidates from micrographs. A typical generative method employs a template-matching technique~\cite{Hall2004, Huang2004} with a cross-correlation similarity measure to accomplish particle selection. The unsupervised approaches distinguish the images of particle-like objects from background noise in micrographs via an unsupervised learning manner (i.e., without any labeled training data)~\cite{Adiga2004, DoG}. The discriminative methods first train a classifier based on a labeled dataset of positive and negative examples, and then apply this trained classifier to detect and recognize particle images from micrographs~\cite{semisupervised, Sorzano2009, PPSVM}. Although these computational approaches have greatly reduced time and effort spent on single-particle data analysis, there still remain gaps to achieve a fully automated pipeline for efficient particle picking. For instance, most of the generative methods require the user to prepare an initial set of high-quality reference particles used as templates to search for similar particle candidates from micrographs, while the discriminative approaches usually demand the user to manually pick a number of positive and negative samples to train the classifier. Thus, these approaches generally depend upon a certain level of human intervention to provide a portion of manually-picked particles to initialize the particle selection process. Although the unsupervised approaches do not heavily rely on hand-labeled data, they rarely fully exploit the intrinsic and unique characters of particles to facilitate automated particle picking. Therefore, the unsupervised approaches are often combined with the template-matching or classification based approaches to achieve decent picking results~\cite{TMaCS, MAPPOS}.

In recent years, deep learning has become an increasing popular tool in the machine learning field probably due to the availability of large-scale training data, the advance of powerful computing platforms and the development of efficient learning algorithms~\cite{Reviewdeep,Deep6,DBLPs}. Tests on several well-known benchmark datasets have demonstrated that deep learning can achieve better performance especially in large-scale data analysis than traditional machine learning approaches~\cite{Reviewdeep,Deep6,ImageNet,DBLP}. So far deep learning has been successfully applied to a wide range of data science fields, such as computer vision~\cite{ImageNet,Facerecognition,Deep2,Deep3}, natural language processing~\cite{Deep6,Deep5,ShowAttendTell} and computational biology~\cite{Deepsea,DeepCom}. More recently, the Google team has shown that the deep learning framework is capable of achieving an impressive and amazing level of artificial intelligence that can closely mimic the problem solving skills of human experts~\cite{Deepgame,Deepgame2}. Despite these successful stories in a variety of applications, it still remains unknown whether the deep learning technique can also be effectively used to address the current problems in particle picking and achieve a fully automated particle selection procedure in single-particle cryo-EM data analysis. Here, we aim to answer this question and propose a deep learning framework, called DeepPicker, to fill the aforementioned gaps in automated particle selection and liberate structural biologists from the dreary manual picking process to focus on more interesting work.

Although our deep learning model can be trained with a semi-automated manner as the conventional classification based approaches, it can be equipped with a new training scheme, which fully exploits the known particles of other molecular complexes that are different from the target one and whose structures have been previously determined by cryo-EM. This new training strategy does not require any manual effort in particle picking for the current target molecular complex and thus is considered \emph{fully automated}. Our deep learning model with such a new cross-molecule training scheme is innovative and can effectively capture the common abstract representation of latent features from the known particles of the previously-determined molecular structures. It can closely mimic human intelligence and effectively use the extracted cross-molecule features to initialize the particle picking process of the current target complex. To our knowledge, our approach is the \emph{first} successful attempt to fully exploit the cross-molecule data to achieve \emph{full} automation in particle picking without any human interference.

We have implemented our deep learning method for particle picking and tested it on the real cryo-EM data of three molecules that have been published in the past three years, including TRPV1~\cite{TRPV1} (from EMPIAR (\url{https://www.ebi.ac.uk/pdbe/emdb/empiar/})), human $\gamma$-secretase~\cite{GAMMA} and yeast spliceosome~\cite{Spliceosome}. Our tests have demonstrated that the proposed deep learning framework with either semi-automated or fully automated training scheme can accurately detect and select a sufficient number of particles that are comparable to those picked manually by human experts. Our new automated picking approach can significantly reduce time and labor spent in single-particle data analysis and thus greatly relieve a bottleneck in the automated cryo-EM structure determination pipeline.

\vspace{-10pt}
\section{Results}

\subsection{A deep learning framework for fully automated particle picking}

\begin{figure}
	\centerline{ \includegraphics[scale=0.4]{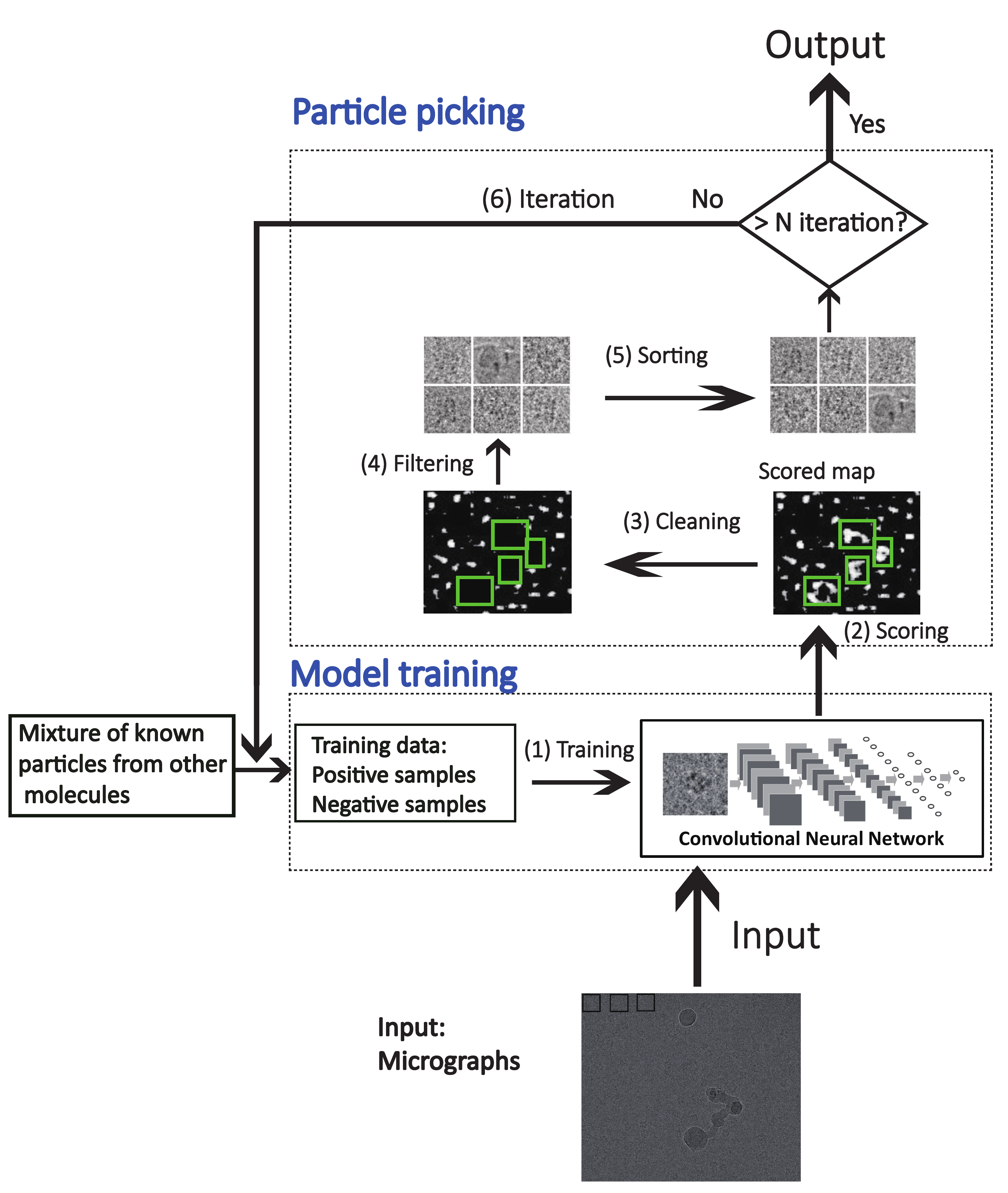}}
	\caption{Schematic overview of our automated particle picking pipeline. The particle selection procedure consists of six steps: (1) training, (2) scoring, (3) cleaning, (4) filtering, (5) sorting and (6) iteration. In our fully automated picking pipeline, the initial training data are obtained from the known particles of other molecular complexes whose structures have been previously solved via cryo-EM.}\label{Methods}
\end{figure}

We implemented a deep learning framework for automated particle picking in single-particle cryo-EM structure determination (Fig.~\ref{Methods}). The automated particle picking pipeline consists of two modules, i.e, model training and particle picking (Fig.~\ref{Methods}). In the model training module, a set of labeled positive and negative samples is used to train a convolutional neural network (CNN) model (see Sections 4.2 and 4.3), while in the particle picking module, the trained CNN classifier is then used to select particle images from input micrographs. The particle picking module is further divided into five steps: scoring, cleaning, filtering, sorting and iteration. In the scoring step, a sliding window (i.e., a square box) of a fixed size is used to scan each micrograph from the top left corner to the bottom right corner with a constant step size. The box size of the sliding window is chosen to be slightly larger than the particle size, which can be easily estimated and defined as a parameter. During the above scanning process, each patch within the sliding window is extracted and fed to the trained CNN classifier as input data. The prediction score between 0 and 1 output by the CNN model, which represents the probability of being a particle at the current position, is then assigned to the center of the corresponding window. After that, we obtain a ``scored map", which describes the distribution of the likelihood scores of particles over the whole micrograph. As ice noise can easily introduce false positives during the picking process, we also employ a cleaning step to discard these false particles from the candidate list. In this cleaning step, we first connect any two neighboring pixels if their prediction scores are both above a threshold, and then examine the size of each connected domain (i.e., the portion of all connected pixels). If the size of a connected domain is larger than a cutoff value, it is regarded as a potential false positive probably due to ice noise, and thus removed from the candidate list. In the filtering step, we aim to refine the current set of particle candidates and also identify the center coordinates of the final remaining particles from the scored map. We first introduce a concept of peak window, the size of which is related to the minimum distance between centers of two possible particles. Then the position with the maximum prediction score in each peak window is chosen and output as the center of a particle. We also remove bad particle candidates in which the number of extreme pixels is more than three standard deviations away from the mean. In the sorting step, we sort the remaining particle candidates according to their prediction scores. In the iteration step, we use the particles picked by the previous CNN classifier which was trained over the known particles of other molecules to further refine the CNN model. After a certain number of iterations, the algorithm outputs the top list of the highest-rated particles. The detailed setting of the parameters described for the above operations is provided in Section 4.5.

In general, the particles of different molecular complexes display distinct sizes and shapes. Thus, how to effectively exploit the cross-molecule features to pick particles from micrographs of the current complex is a major challenge in our work. As described above, this problem has been tackled in our framework mainly using two strategies. First, the known particles from multiple types of other molecular complexes are combined together to train the deep learning classifier and enable it to capture common features of particles. Second, the particles of the target molecule identified from the deep learning classifier trained by a cross-molecule manner in the first iteration are further used to refine the model to better describe the features of the current target complex. In our automated particle picking scheme, the user is not required to manually pick any particle from micrographs of the current target molecule. Thus, it is considered a fully automated procedure without requiring any manual intervention during the particle selection process.

\subsection{Performance evaluation on fully automated particle picking}

\subsubsection{Accuracy of particle picking}

\begin{figure}[ht]
\centering
\subfigure[]{\includegraphics[width=5cm,height=4.84cm]{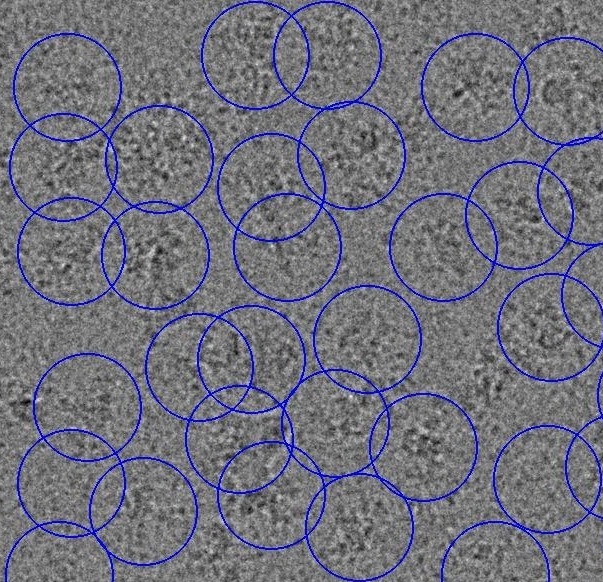}}
\hspace{2ex}
\subfigure[]{\includegraphics[width=8cm,height=4.7cm]{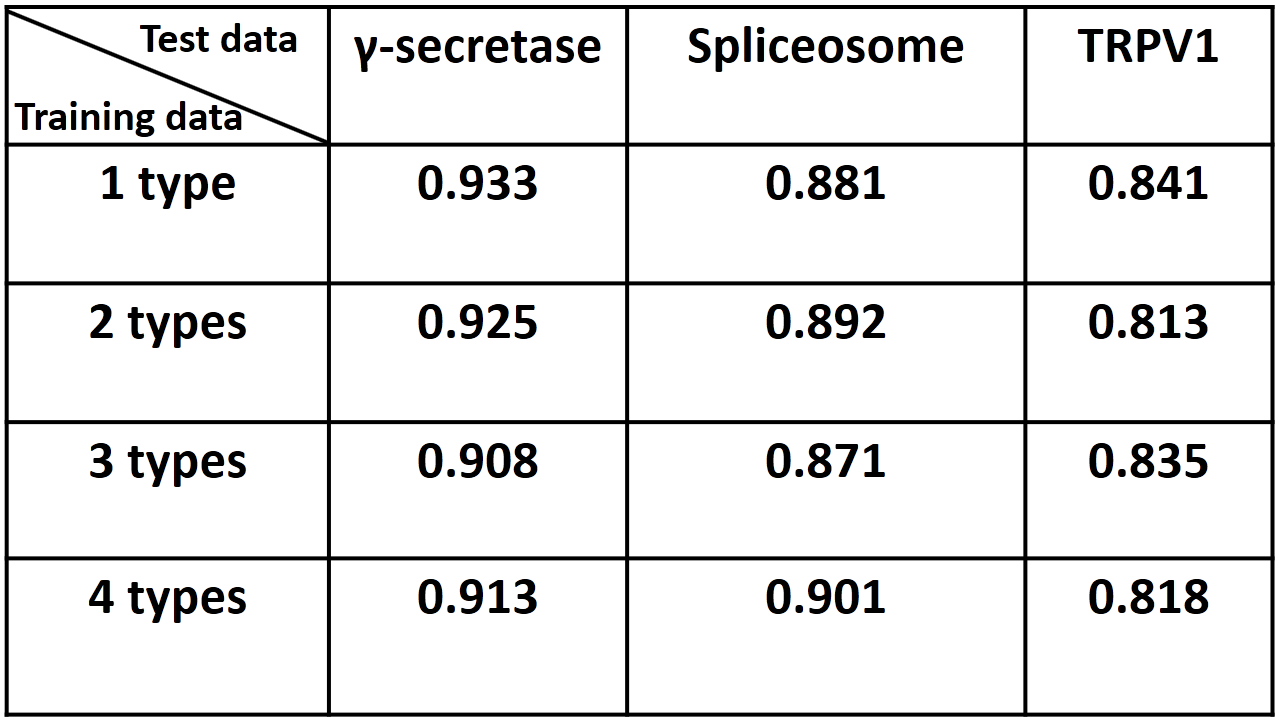}}
\subfigure[]{\includegraphics[width=6cm,height=4.6cm]{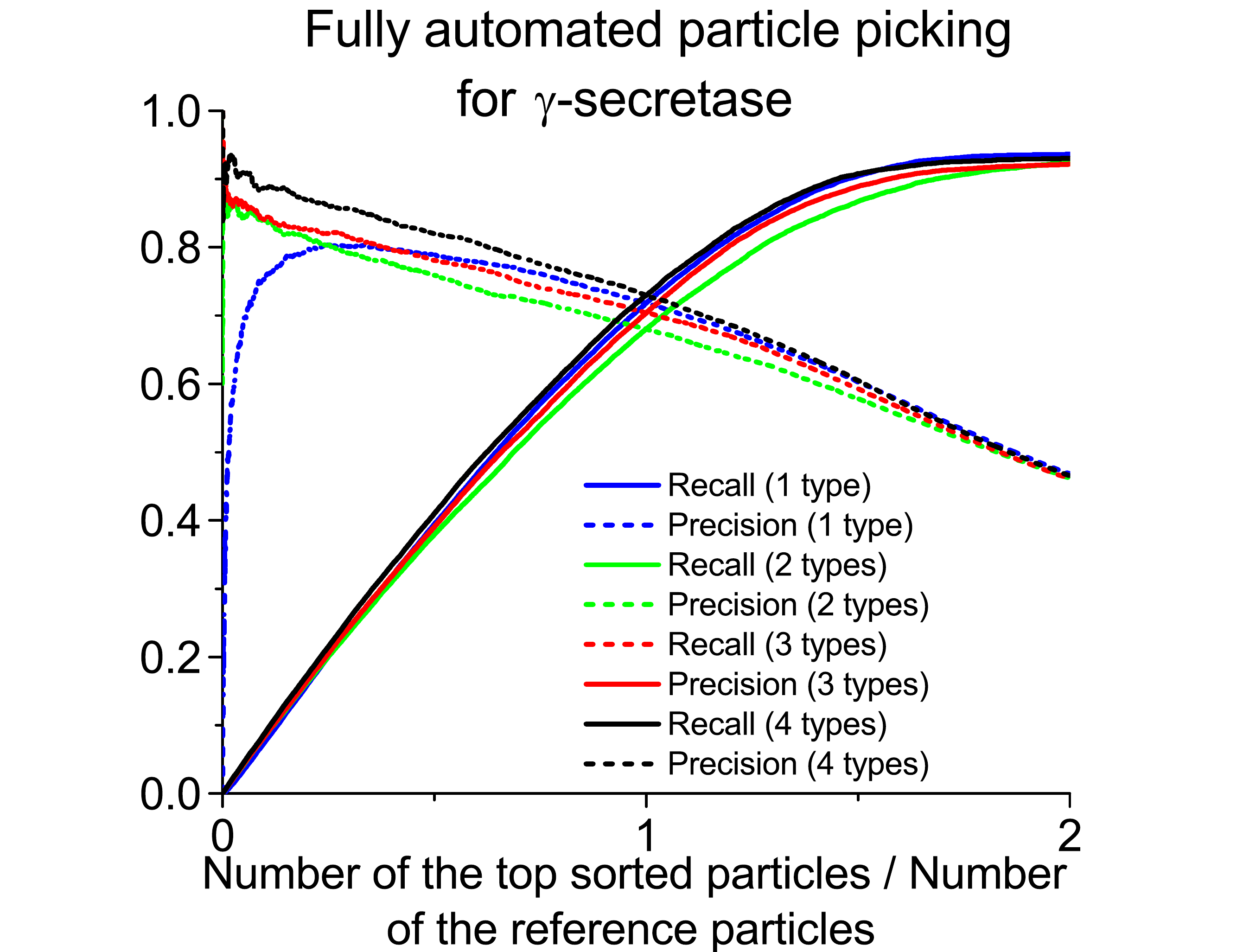}}
\hspace{-5ex}
\subfigure[]{\includegraphics[width=6cm,height=4.6cm]{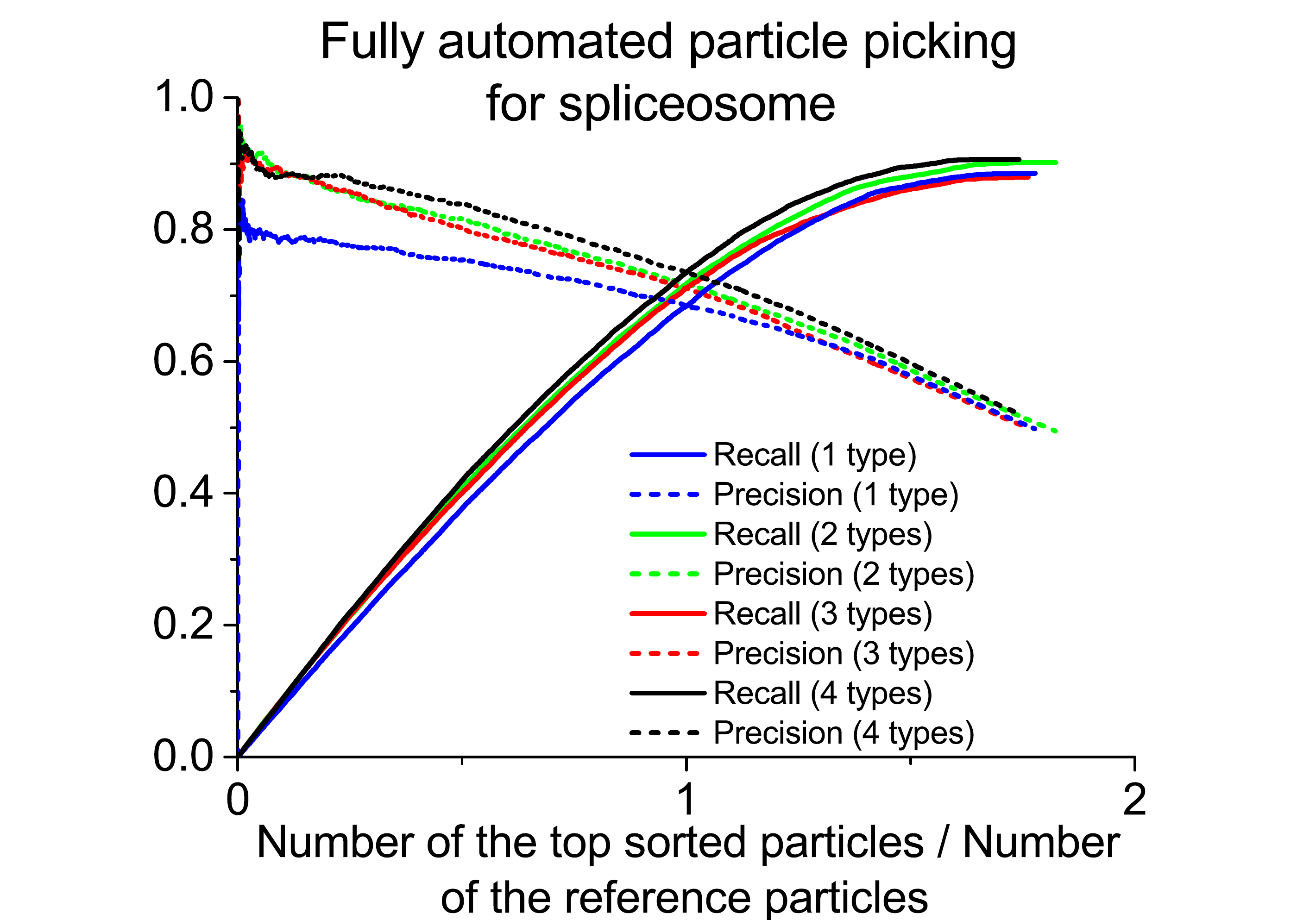}}
\hspace{-5ex}
\subfigure[]{\includegraphics[width=6cm,height=4.6cm]{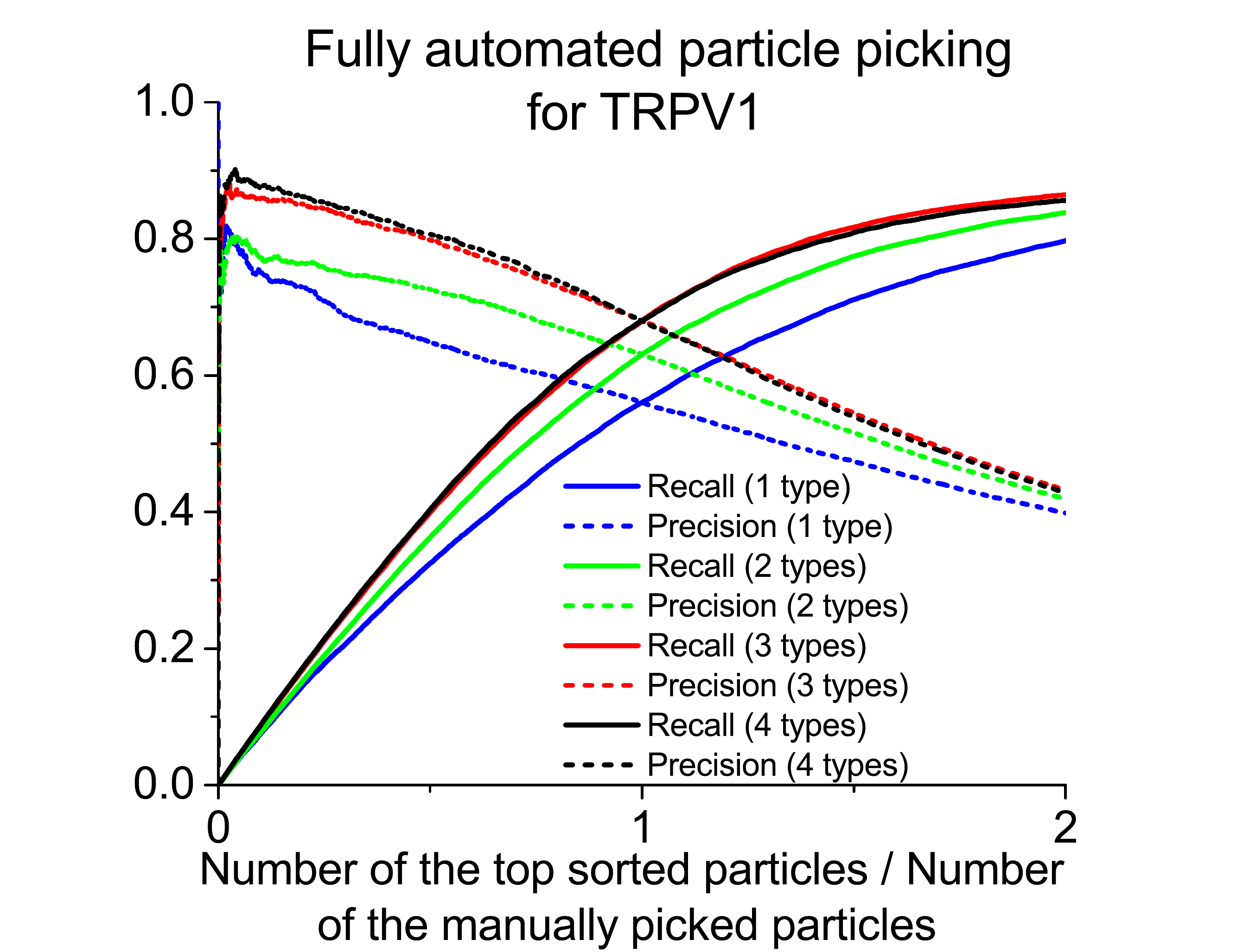}}
\caption{Results on fully automated particle picking. (a) An example of our fully automated particle picking results for spliceosome, using a mixture of 5000 TRPV1 and 5000 $\gamma$-secretase particles picked manually by human experts as training data. (b) The summary of recall scores for $\gamma$-secretase, TRPV1 and spliceosome using different combinations of training data. The recall scores were computed on those picked particles with prediction scores above 0.5. (c), (d) and (e) The trends of precision and recall vs. the number of picked particles for $\gamma$-secretase, spliceosome and TRPV1, respectively. The horizontal axis represents the ratio of the number of the top sorted particles picked by our fully automated method vs. the number of manually picked particles in the reference set. The blue, green, red and black lines represent the results of using 10000 manually picked particles as training data contributed roughly equally by one, two, three and four types of other molecules (which were different from the target molecule), respectively. The real and dashed lines represent the recall and precision scores, respectively. More details about different combinations of known particles from different molecules in training data can be found in Section 4.3.}\label{AutoPR}
\end{figure}

We have tested and evaluated the performance of our fully automated particle picking approach on the published datasets of $\gamma$-secretase, spliceosome and TRPV1 (see Section 4.1 for details of data preprocessing), in which the known particles of the other two molecules that were different from the target complex were used as training data (Fig.~\ref{AutoPR}). In addition, another two datasets, including $\beta$-galactosidase~\cite{Beta} and NSF (N-ethylmaleimide Sensitive Factor) fusion complex (20S particle)~\cite{SNARE}, were incorporated into training data. We compared the fully automated picking results with the coordinates of the reference particles identified manually by human experts and measured both precision and recall scores (see Section 4.4). As in the current single-particle analysis pipeline, the fraction of true particles that have been picked is generally more important for 3D model reconstruction~\cite{Langlois20141}, we focused more on the recall metric.

A typical example of the fully automated picking results is shown in Fig.~\ref{AutoPR} (a), which has demonstrated that most of the particles picked by our fully automated approach from a micrograph were reasonable. All different tests (Fig.~\ref{AutoPR} (b)) show that our fully automated picking method can achieve high recall scores (above 0.81), which implies that most of the reference particles manually picked by human experts can also be identified by our deep learning framework with a cross-molecule training strategy. In addition, when the number of the top scoring particles picked by our fully automated approach was close to that of the reference particles, both recall and precision scores were relatively high (Fig.~\ref{AutoPR} (c), (d) and (e)). These results imply that a major fraction of the particles identified by our fully automated scheme were consistent with those manually picked by human experts. The comparisons among results by using different training datasets (see Section 4.3) show that our CNN classifier trained by multiple datasets achieved more robust performance than that using only a single dataset as training data (Fig.~\ref{AutoPR} (c), (d) and (e)). Such a result was expected, as the hidden features derived from multiple datasets were supposed to be more general. In practice, we can include as many known datasets as possible into training data to boost the classification performance of our deep learning model.

All the above results indicate that, despite different contrast levels of micrographs, different shapes and sizes of the known particles from other molecules used as training data, our fully automated approach can effectively identified accurate particles that were comparable to the manual picking results and thus achieve a near human level of particle selection. Therefore, our fully automated particle picking method can be practically useful for cryo-EM structure determination. As it does not require any human intervention, in practice, it can help the user save a huge amount of time and effort in single-particle data analysis.

\subsubsection{The 2D clustering and class averaging results}

\begin{figure}[ht]
\centering
\subfigure[]{\includegraphics[width=5.4cm,height=4.5cm]{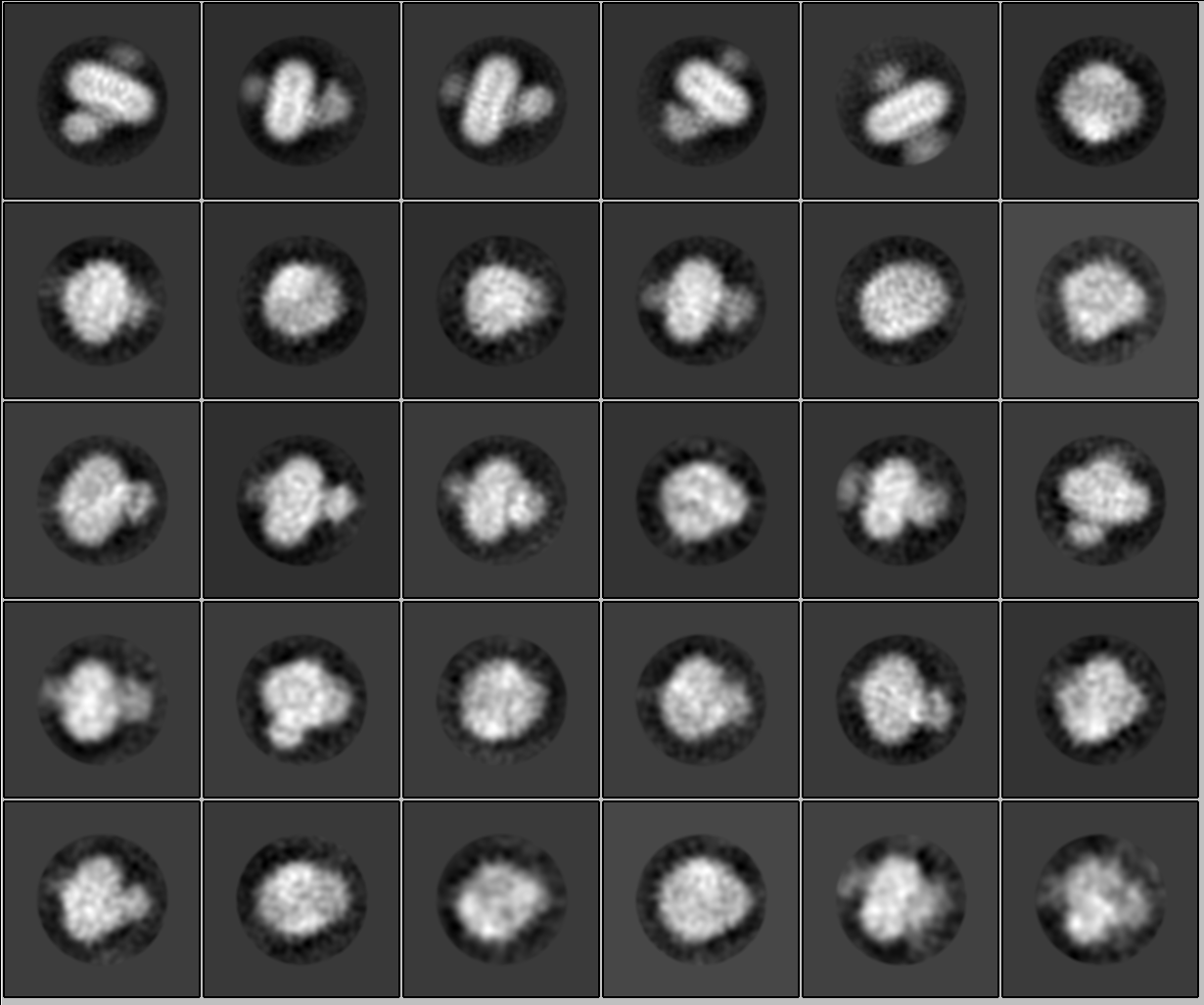}}
\hspace{2ex}
\subfigure[]{\includegraphics[width=5.4cm,height=4.5cm]{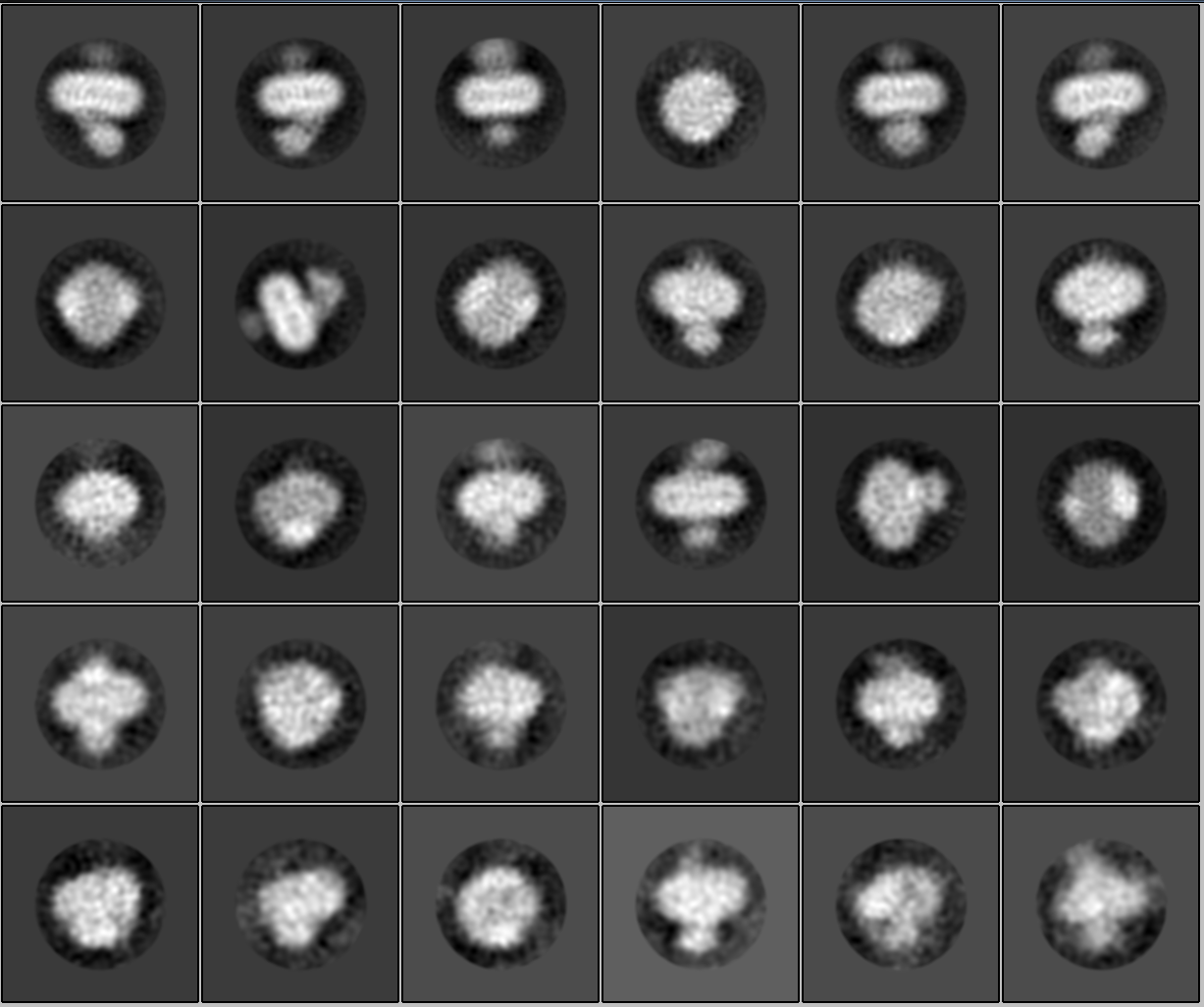}}
\subfigure[]{\includegraphics[width=5.4cm,height=4.5cm]{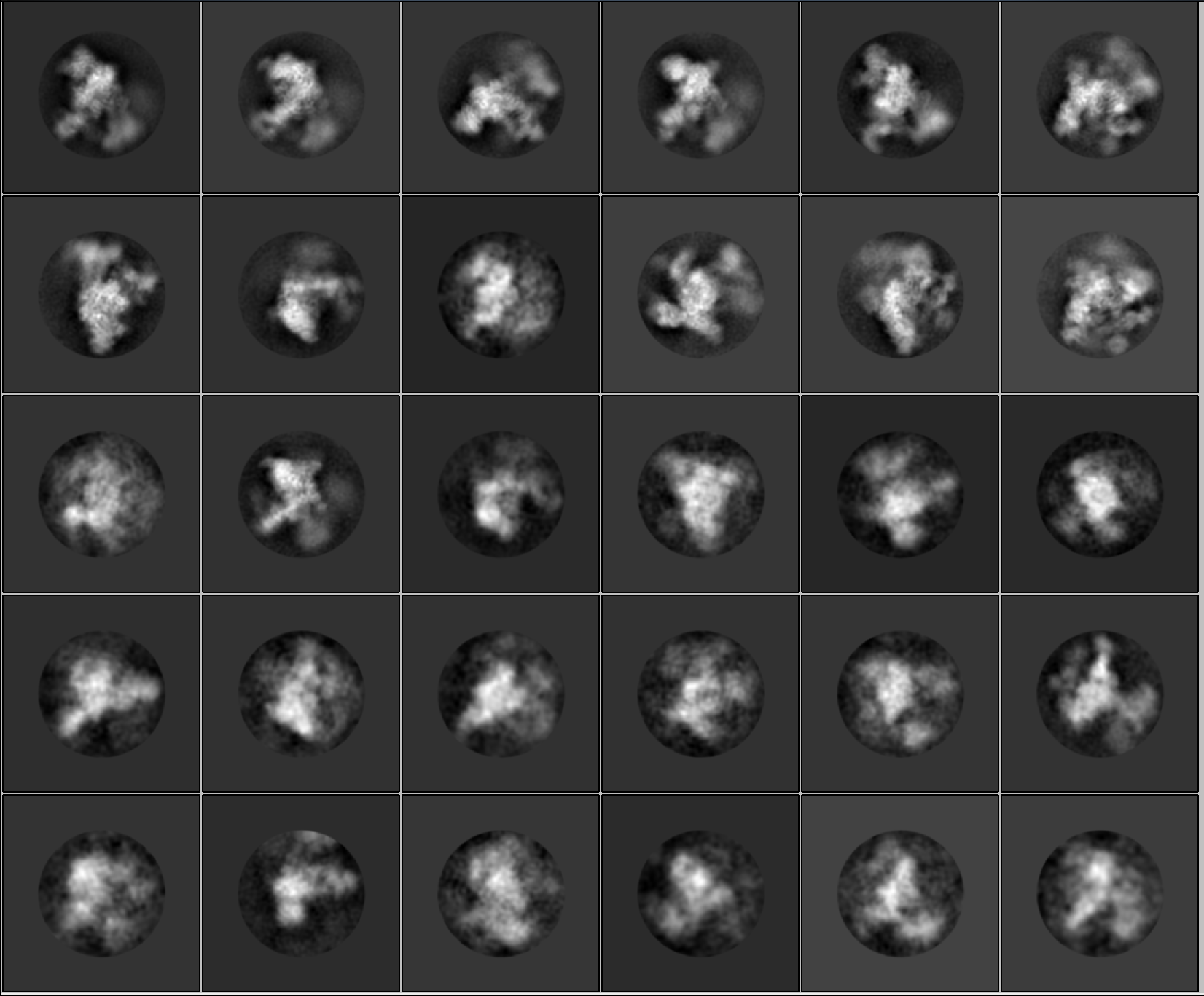}}
\hspace{2ex}
\subfigure[]{\includegraphics[width=5.4cm,height=4.5cm]{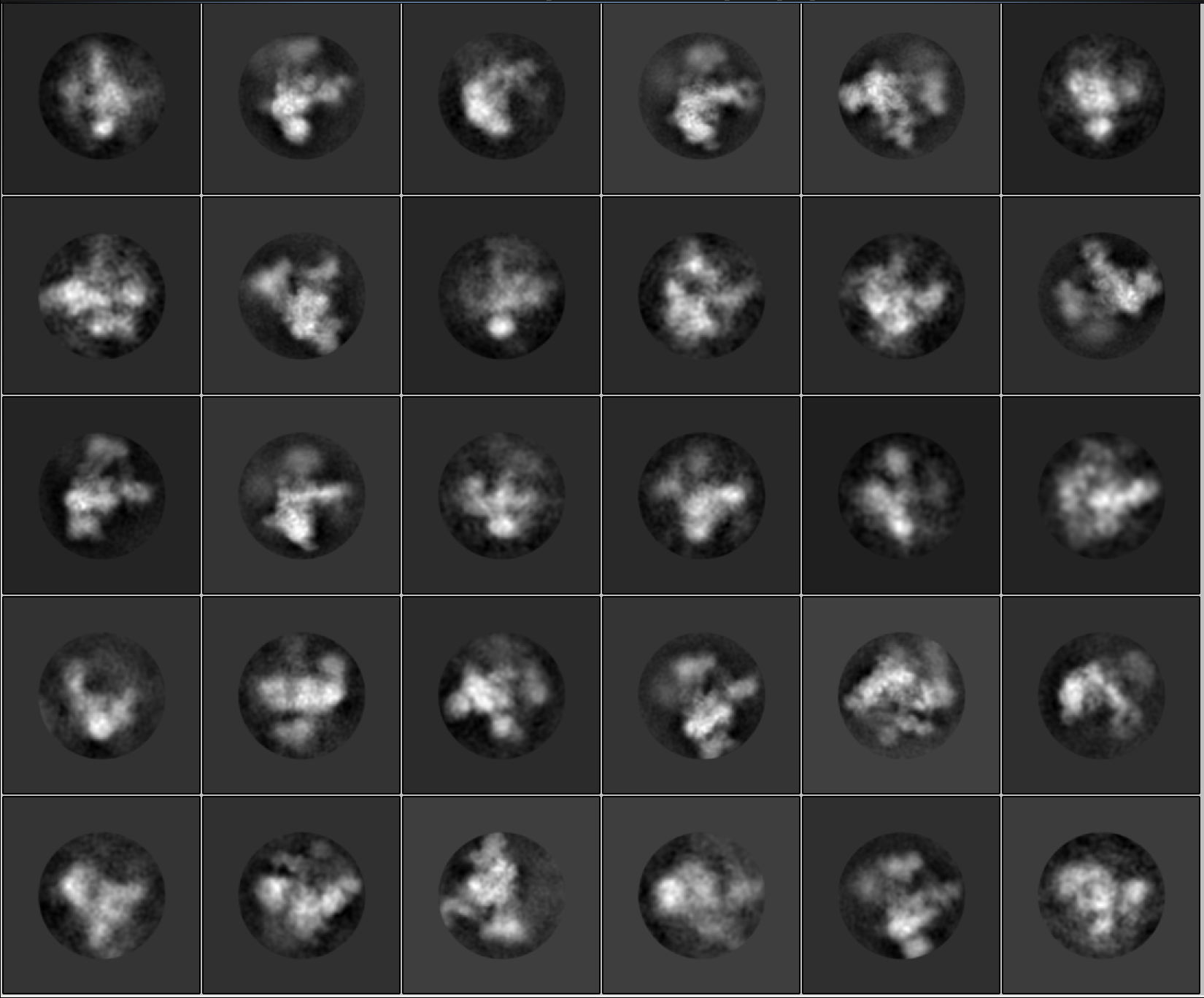}}
\subfigure[]{\includegraphics[width=5.4cm,height=4.5cm]{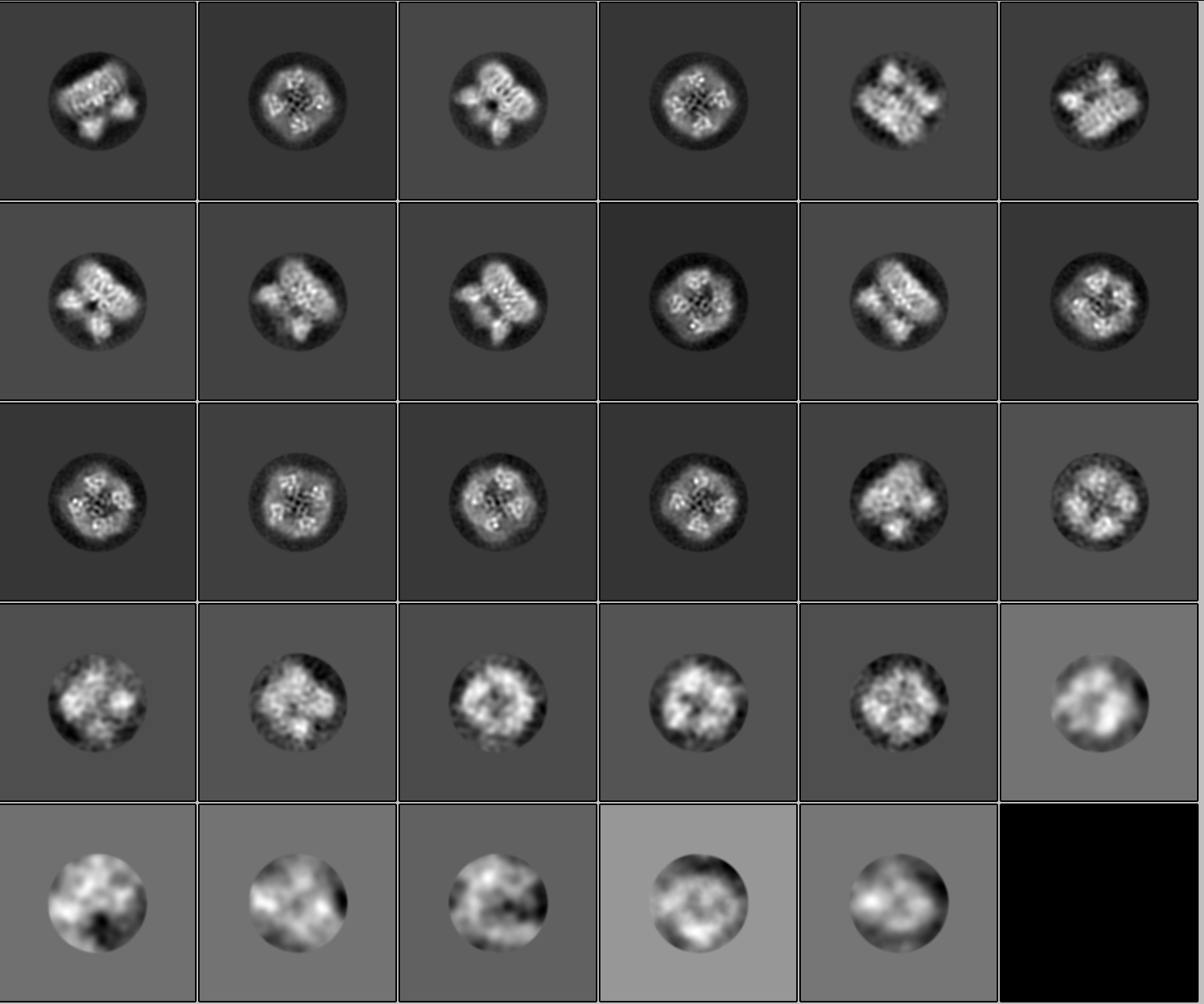}}
\hspace{2ex}
\subfigure[]{\includegraphics[width=5.4cm,height=4.5cm]{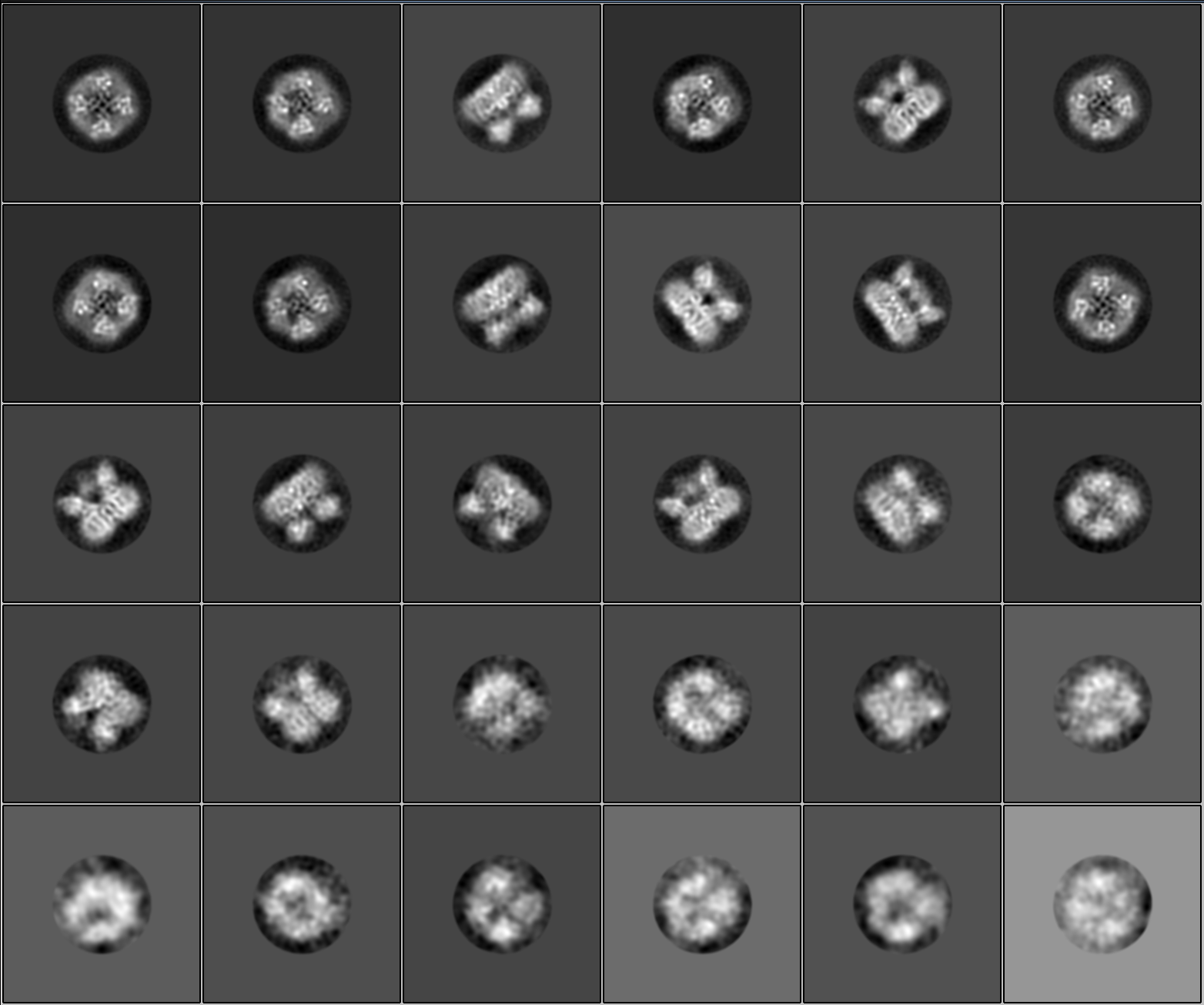}}
\caption{The comparisons between the 2D clustering and class averaging results derived from the particles picked manually by human experts and fully automatically by our deep learning framework. (a), (c) and (e) The 2D clustering and class averaging results of the reference particles picked manually by human experts for $\gamma$-secretase, spliceosome and TRPV1, respectively. (b), (d) and (f) The 2D clustering and class averaging results of the particles picked fully automatically by our CNN classifier for $\gamma$-secretase, spliceosome and TRPV1, respectively. A mixture of 10000 known particles of the other two molecules (each contributing to 5000 particles) among $\gamma$-secretase, spliceosome and TRPV1, which were different from the target molecule, were used as training data. All 30 classes are shown.}\label{2D}
\end{figure}

The 2D clustering and class averaging operations are commonly used right after the particle picking step in the single-particle data analysis pipeline to further remove false positives and refine the list of the selected particles for 3D map reconstruction~\cite{2D,relion}. Thus, an additional effective method for evaluating the practicability of an automated particle picking approach is to further examine the 2D clustering and class averaging results of the identified particles. We have performed such an investigation and refined the original list of the picked particles by filtering those obviously bad-quality particles (see Section 4.4) during the 2D clustering and averaging operations as in a standard single-particle data analysis pipeline. The comparison results have demonstrated that the 2D clustering and class averaging results of our fully automated picking method were comparable to those of the reference particles manually selected by human experts (Fig.~\ref{2D}). For example, the side-view particles of $\gamma$-secretase that were identified and used in the original paper~\cite{GAMMA} to reconstruct the near-atomic resolution 3D map were almost all present in our automated results (Fig.~\ref{2D} (a) and (b)). This implies that the particles automatically picked by our approach can represent a majority of good-quality 2D images that were practically useful for downstream data analysis in cryo-EM structure determination. Thus, our fully automated particle picking approach can provide a good starting point for 3D map reconstruction in cryo-EM.

\subsubsection{Identification of particle centers}

\begin{figure}[ht]
\centering
\subfigure[]{\includegraphics[width=4cm,height=3cm]{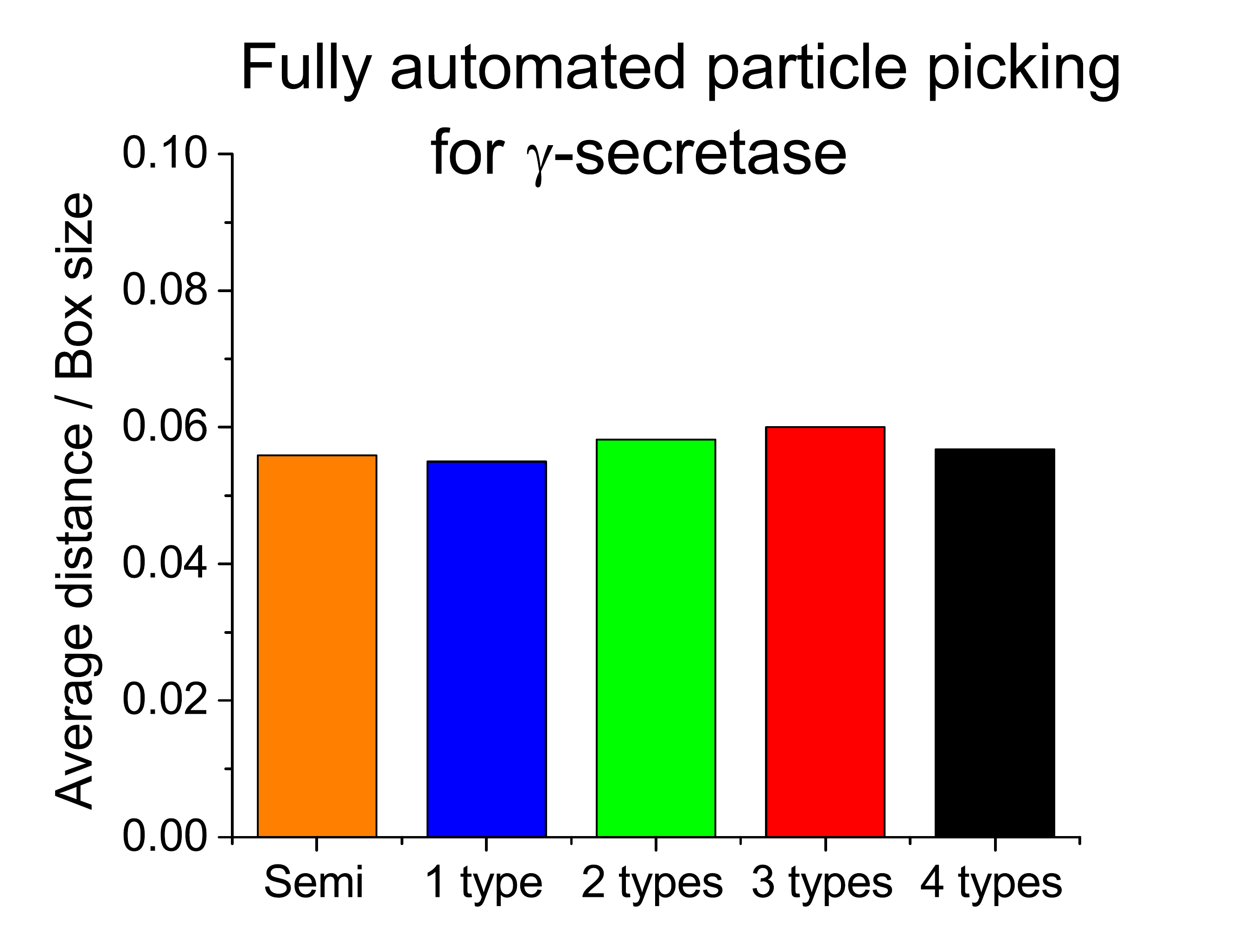}}
\subfigure[]{\includegraphics[width=4cm,height=3cm]{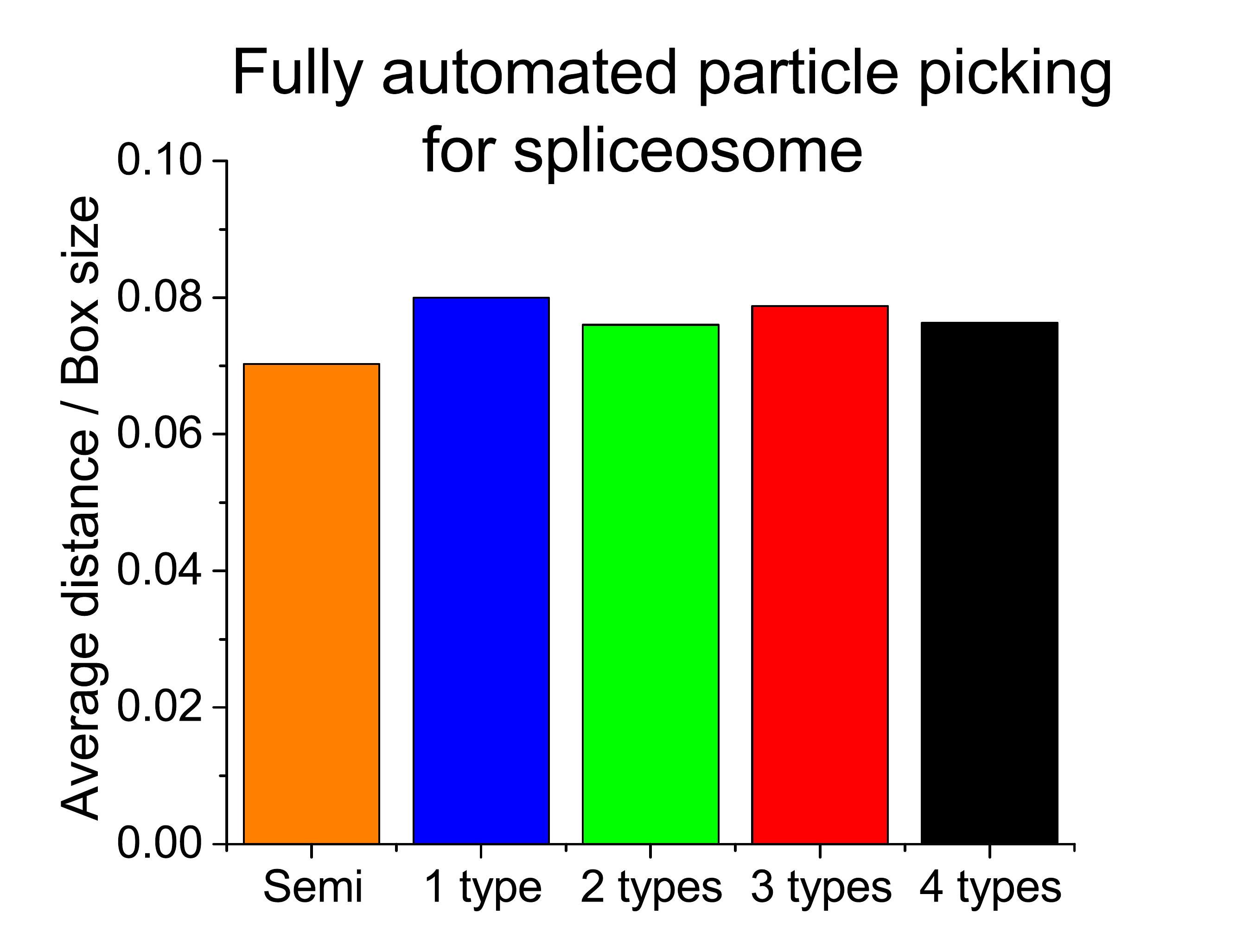}}
\subfigure[]{\includegraphics[width=4cm,height=3cm]{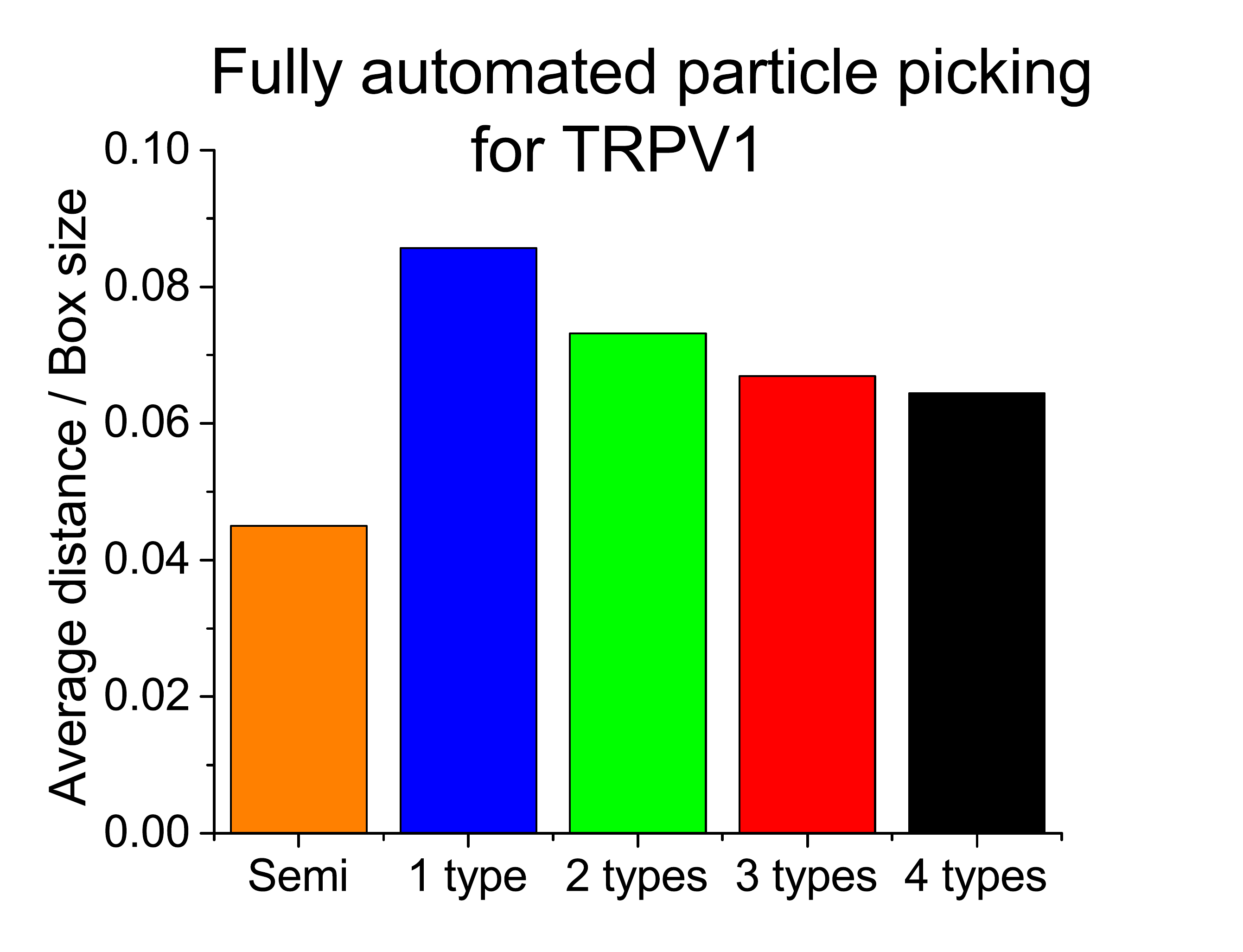}}
\subfigure[]{\includegraphics[width=6cm,height=6cm]{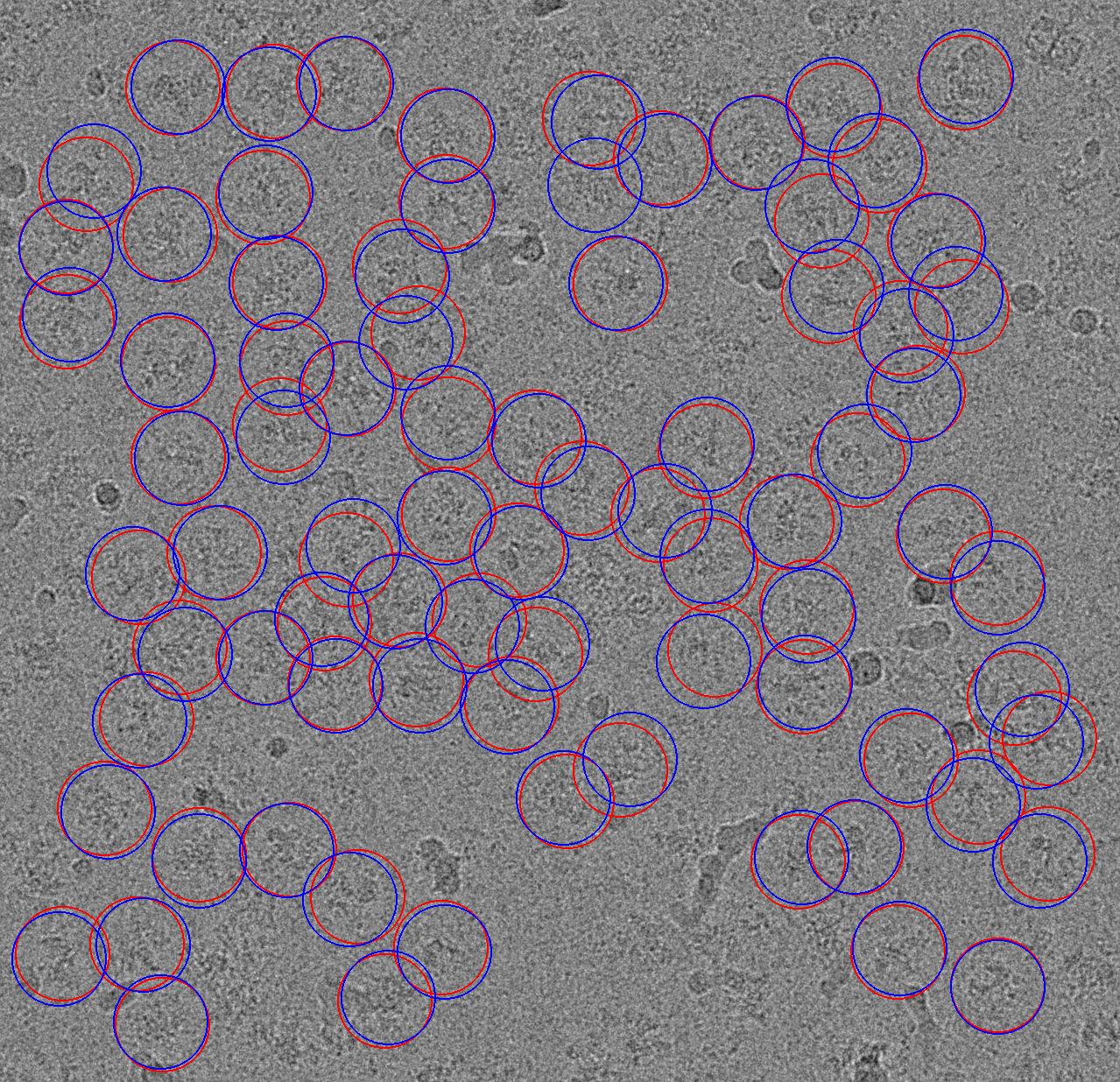}}
\caption{Results on the identifying the centers of particles. (a), (b) and (c) The average distance between our automatically picked particles and their manually picked counterparts in the reference set, normalized by the box size of sliding window for $\gamma$-secretase, spliceosome and TRPV1, respectively. The orange bars represent the results of the CNN classifier trained by 10000 particles of the same molecule (such a scheme is also called semi-automated picking, see also Section 2.3). The blue, green, red and black bars represent the results of the CNN classifier trained by a mixture of 10000 known particles roughly equally contributed by one, two, three and four types of other molecules that were different from the target complex, respectively. (d) An example of the comparison between the spliceosome particles picked manually by human experts and fully automatically by our CNN classifier, which was trained by a mixture of 5000 $\gamma$-secretase and 5000 TRPV1 particles. The red and blue circles represent the particles selected by human experts and by our fully automated approach, respectively.}\label{Distance}
\end{figure}

Identifying the accurate centers of particles is crucial for downstream single-particle data analysis, such as 2D clustering and 3D map reconstruction. In the previous 2D clustering and averaging results, we have shown that most of the particles picked by our fully automated approach were well centralized: almost all the averaging particle images selected by our program were fully covered by the mask circle (Fig.~\ref{2D}). To further investigate the accuracy of our fully automated approach in identifying the centers of particles, we also measured the average distance with respect to the box size of sliding window between the particles picked fully automatically by our CNN classifier and manually by human experts (Fig.~\ref{Distance}). We observed that the particles picked by our fully automated scheme deviated less than 10\% of the box size from the reference images. These results imply that our fully automated approach can accurately detect the centers of particles that were quite close to those manually identified by human experts.

\subsection{Semi-automated particle picking with an alternative training strategy}

\begin{figure}[ht]
\centering
\subfigure[]{\includegraphics[width=7cm,height=4.2cm]{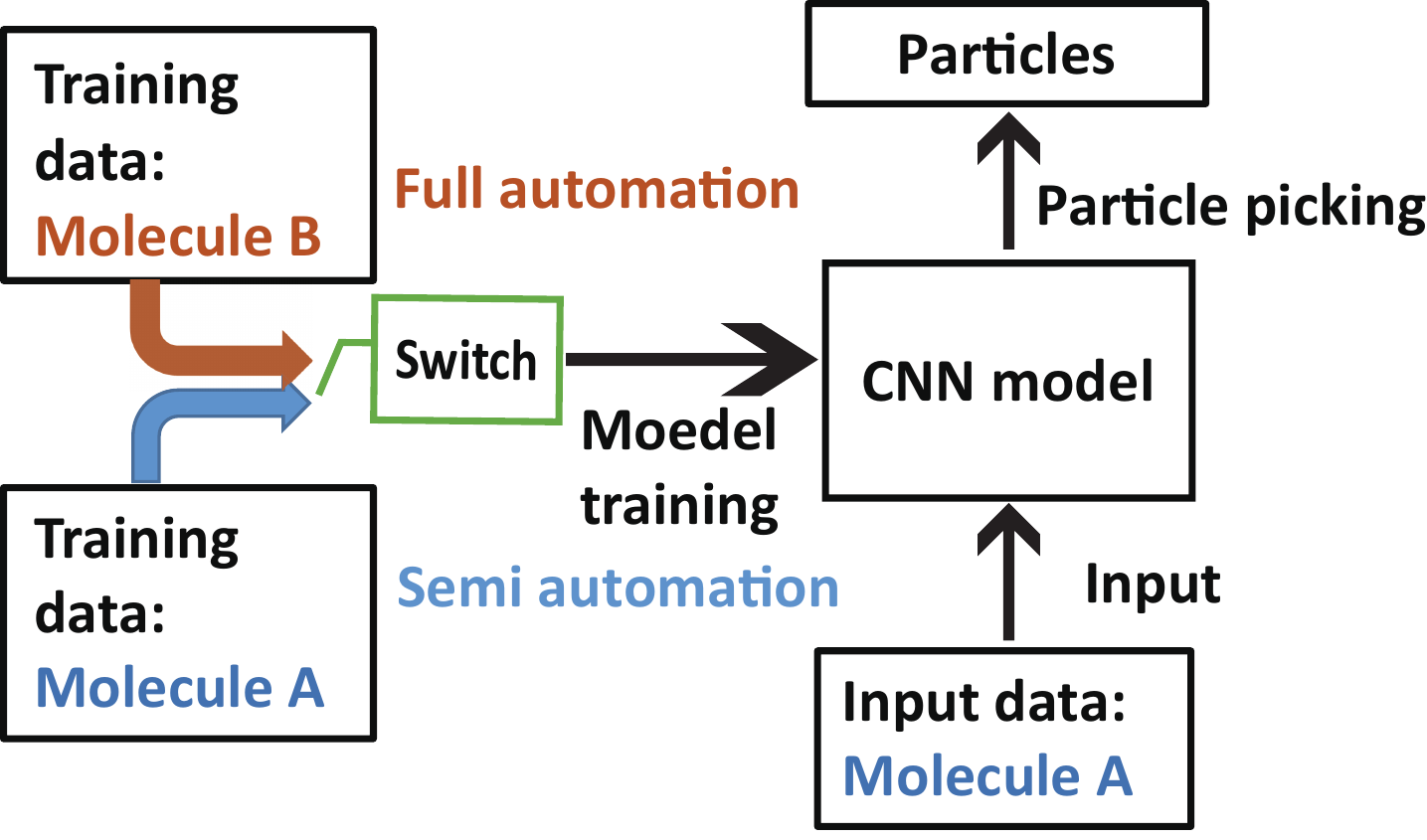}}
\hspace{2ex}
\subfigure[]{\includegraphics[width=8cm,height=4.7cm]{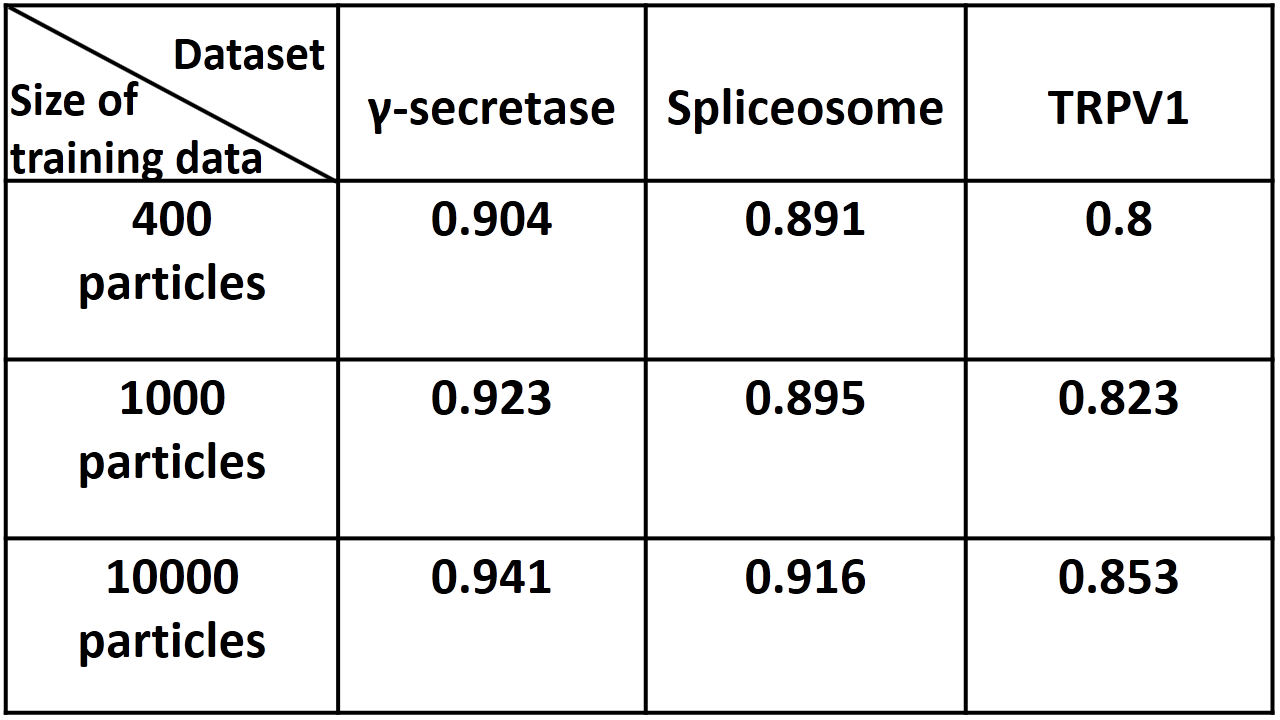}}
\subfigure[]{\includegraphics[width=6cm,height=4.6cm]{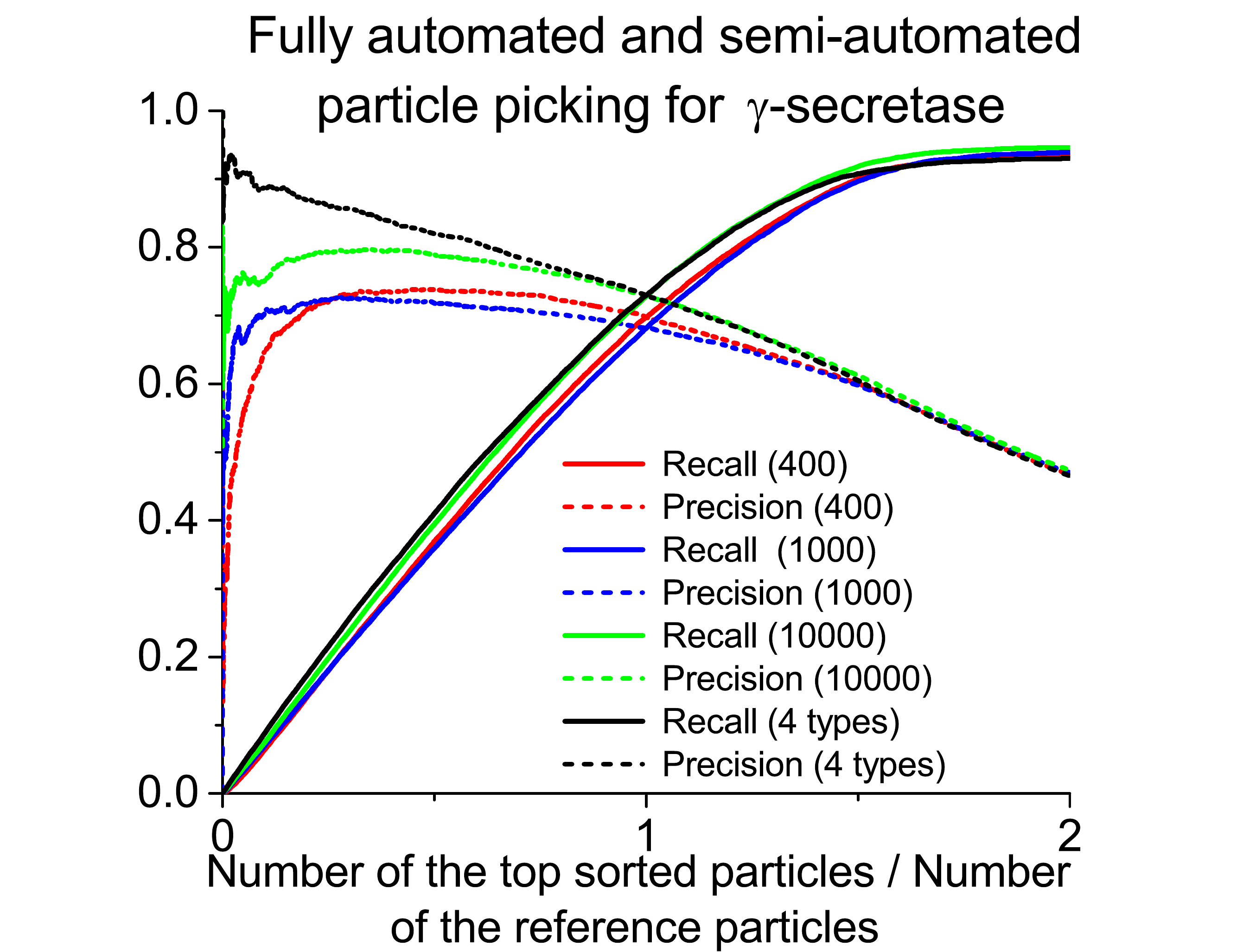}}
\hspace{-5ex}
\subfigure[]{\includegraphics[width=6cm,height=4.6cm]{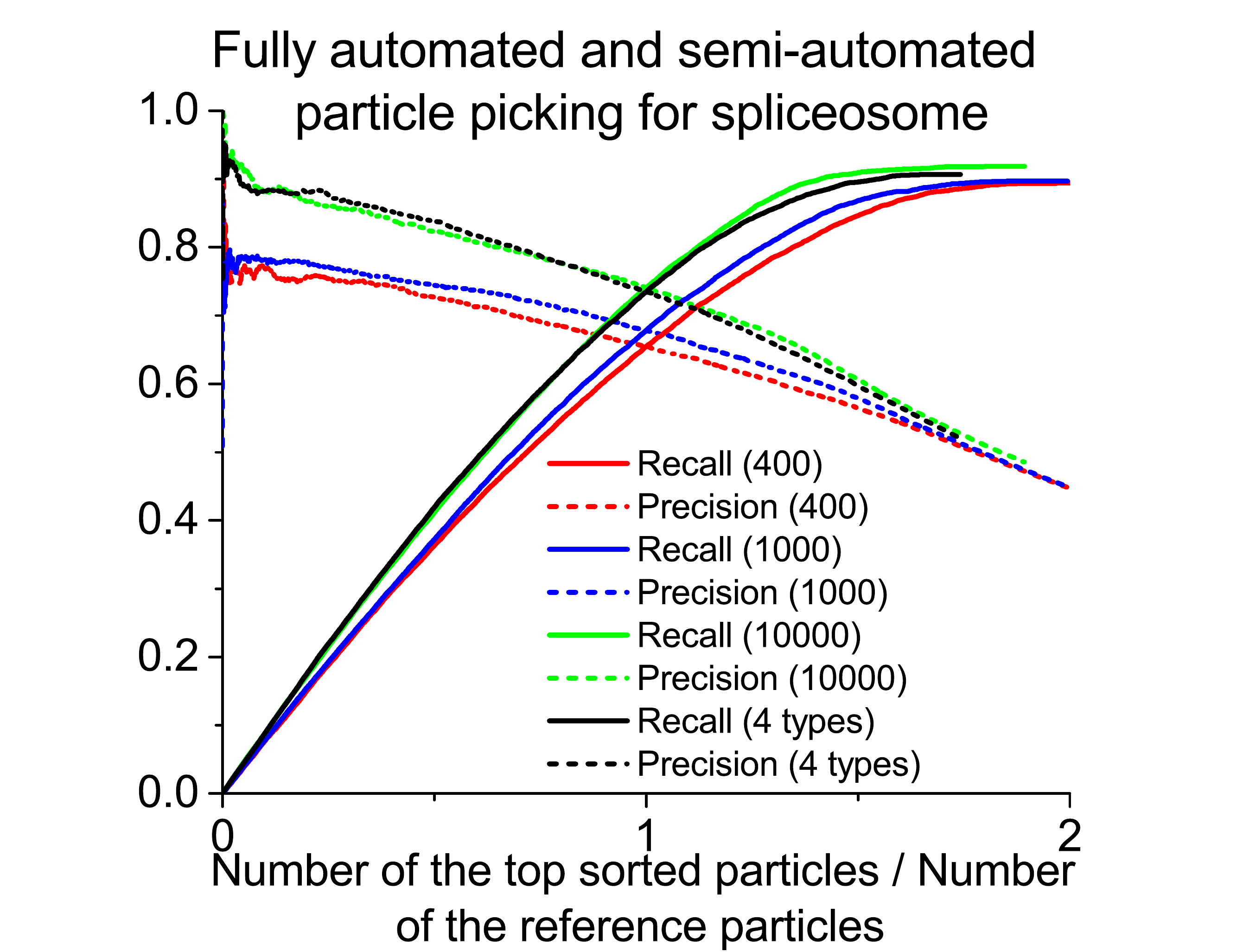}}
\hspace{-5ex}
\subfigure[]{\includegraphics[width=6cm,height=4.6cm]{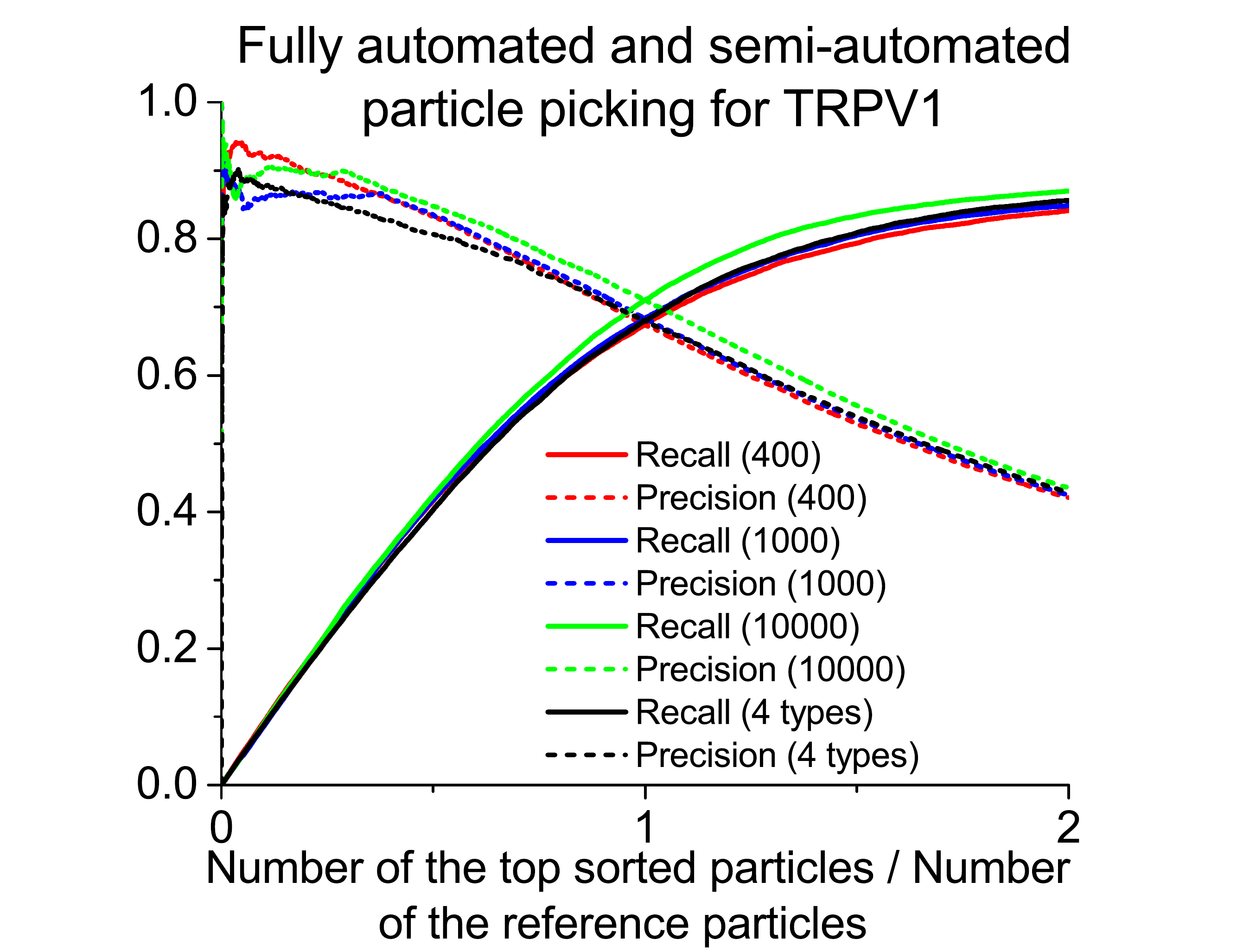}}
\caption{Results on semi-automated particle picking of our deep learning framework with an alternative training scheme. (a) A simple illustration on the alternative choice between fully automated and semi-automated training schemes. (b) The summary of recall scores with prediction scores above 0.5 for the semi-automated particle picking scheme under our deep learning framework for three different datasets with various sizes of training data. (c), (d) and (e) The trends of precision and recall vs. the number of the picked particles for $\gamma$-secretase, spliceosome and TRPV1, respectively. The horizontal axis represents the ratio between the number of the top sorted particles picked by both semi-automated and fully automated schemes vs. the number of manually picked particles in the reference set. The red, blue and green lines represent the semi-automated results of the CNN classifier trained by 400, 1000 and 10000 particles (which were manually picked from micrographs of the target molecule), respectively. The black lines represent the fully automated results in which a mixture of the manually picked particles of four other molecules that were different from the target molecule were used as training data, each contributing to 2500 particles. The real and dashed lines represent the recall and precision scores, respectively.}\label{SemiPR}
\end{figure}

In our fully automated picking scheme, the known particles from other molecular complexes that are different from the current target molecule are used to train the deep learning classifier. An alternative training scheme is to let the user manually select a small number of particles as positive samples to train the CNN model and initialize the particle selection process (Fig.~\ref{SemiPR} (a)). Such a scheme requires a certain level of human intervention and thus is considered semi-automated. As in the previous evaluation of fully automated particle picking, we also performed the similar tests for the semi-automated picking scheme on the same datasets of $\gamma$-secretase, spliceosome and TRPV1.

Overall, various tests on these three datasets have demonstrated that the semi-automated picking method resulting from the alternative training scheme can detect and identify a high percentage of correct particles that were consistent with the manual picking results by human experts (Fig.~\ref{SemiPR} (b)). These results were similar to those in the previous tests of fully automated picking (Fig.~\ref{AutoPR} (b)). The semi-automated picking results with 10,000 particles as training data were only slightly better than the fully automated picking results (Fig.~\ref{SemiPR} (c), (d) and (e)). We also observed similar trends of precision and recall curves with respect to the number of picked particles for both fully automated and semi-automated schemes (Fig.~\ref{SemiPR} (b), (c), (d) and (e)). From the other perspective, our fully automated picking scheme can achieve performance close to that of semi-automated picking, and both schemes can yield outcomes that were comparable to those manual picking results by human experts.

Another noticeable observation in the test results of semi-automated particle picking is that, when the number of manually picked particles used in training data varied from 400 to 10000, the performance of our semi-automated picking method did not significantly change (Fig.~\ref{SemiPR} (b), (c), (d) and (e)). This result indicates that in practice the user only needs to select a small number (e.g., 400) of particles, which is sufficient enough to train our deep learning model for accurately picking new particles from the remaining micrographs. Thus, it can only demand a minimum of human intervention and can also provide a practically useful tool for the current cryo-EM structure determination pipeline.

\vspace{-10pt}
\section{Discussion}

\begin{figure}[ht]
\centering
\subfigure[]{\includegraphics[width=4cm,height=3.8cm]{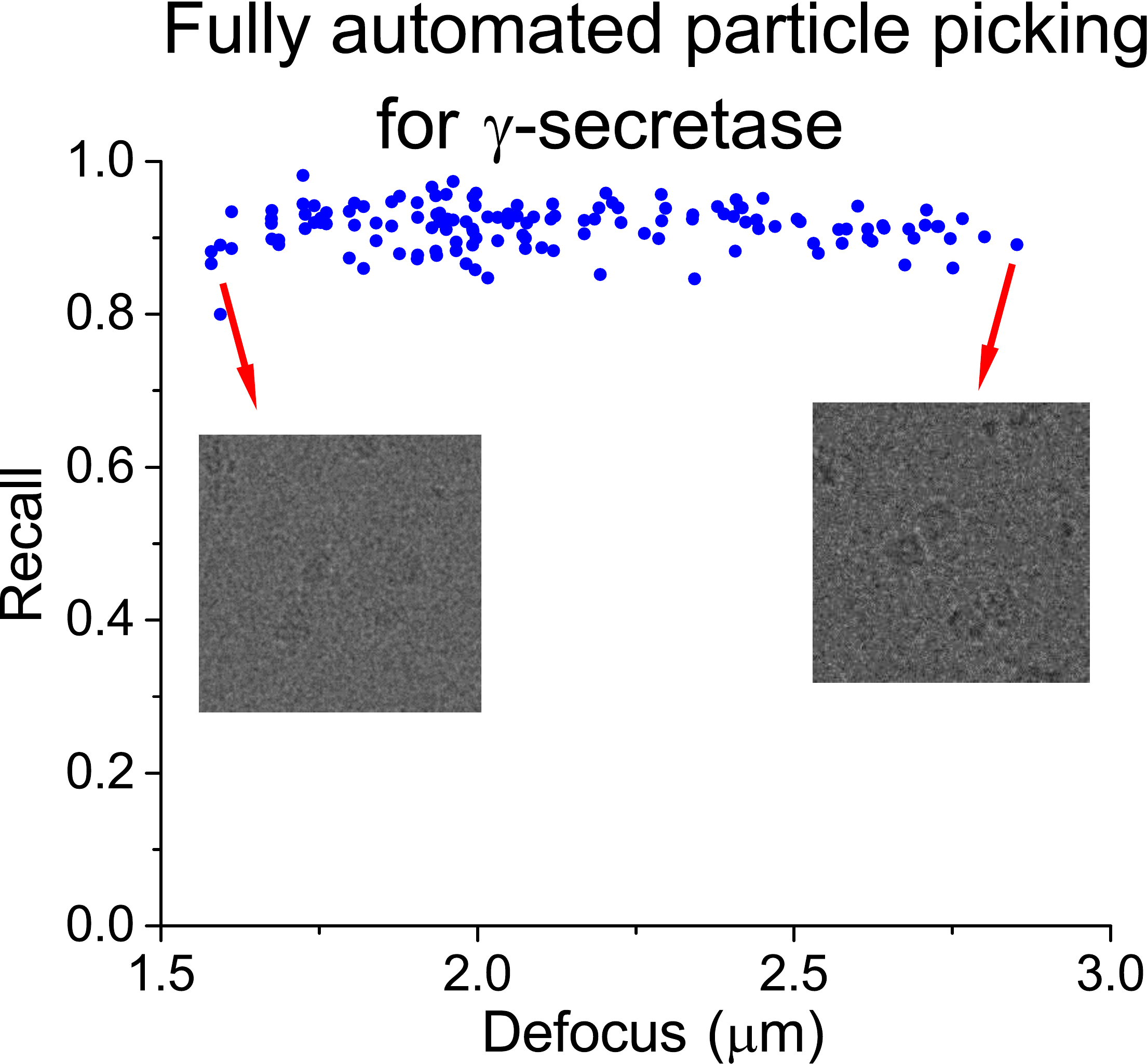}}
\subfigure[]{\includegraphics[width=4cm,height=3.8cm]{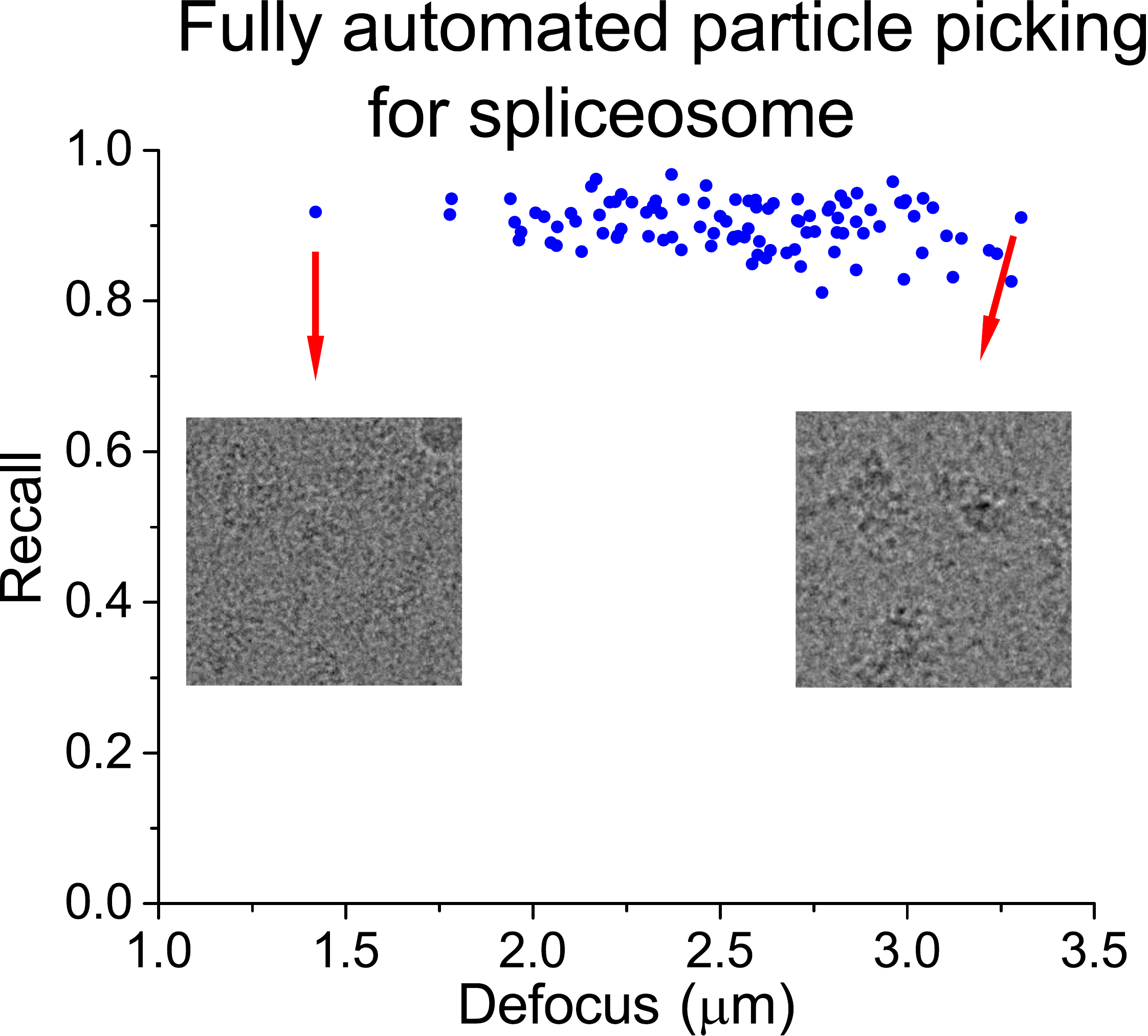}}
\subfigure[]{\includegraphics[width=4cm,height=3.8cm]{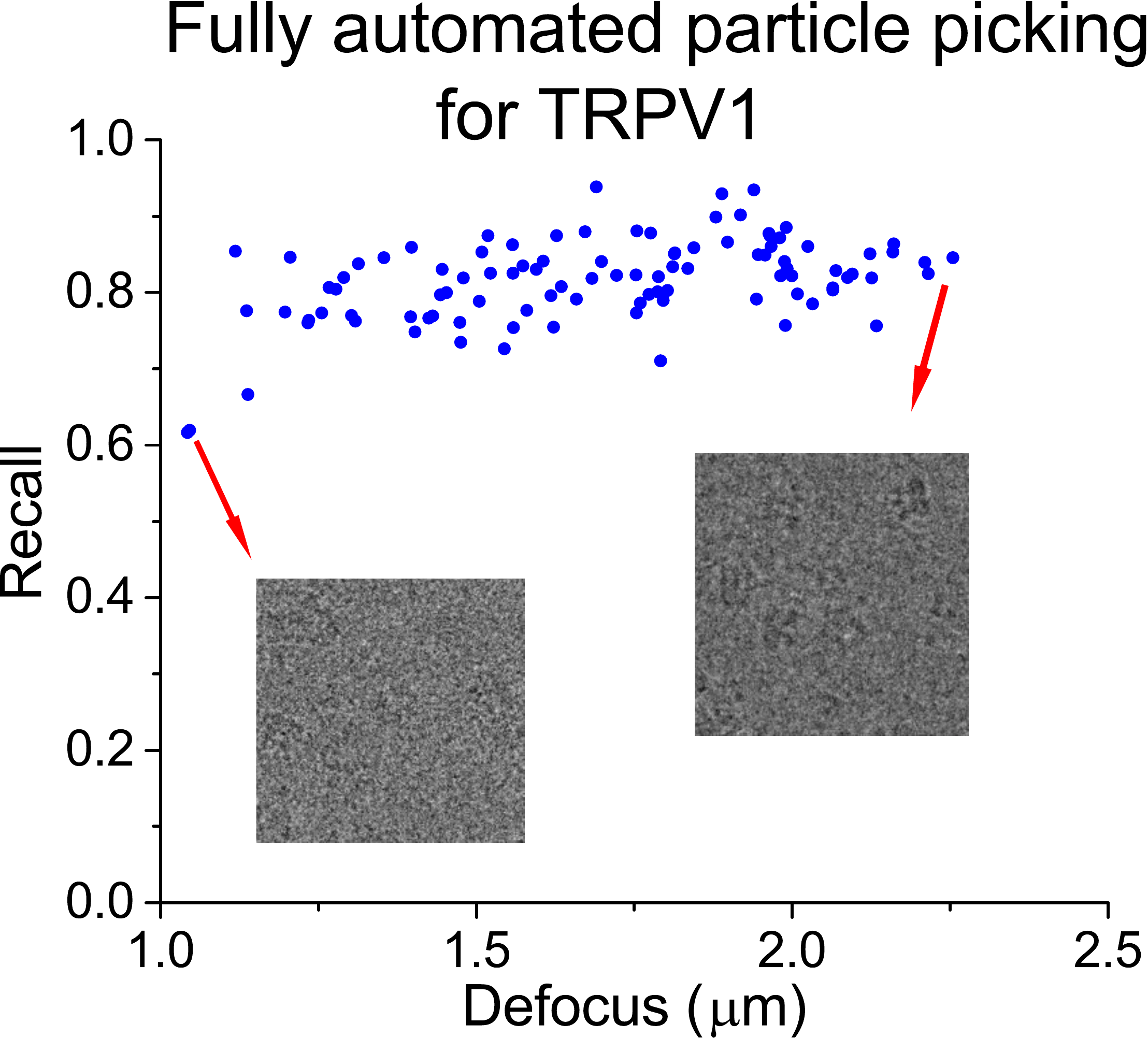}}
\subfigure[]{\includegraphics[width=5.2cm,height=4cm]{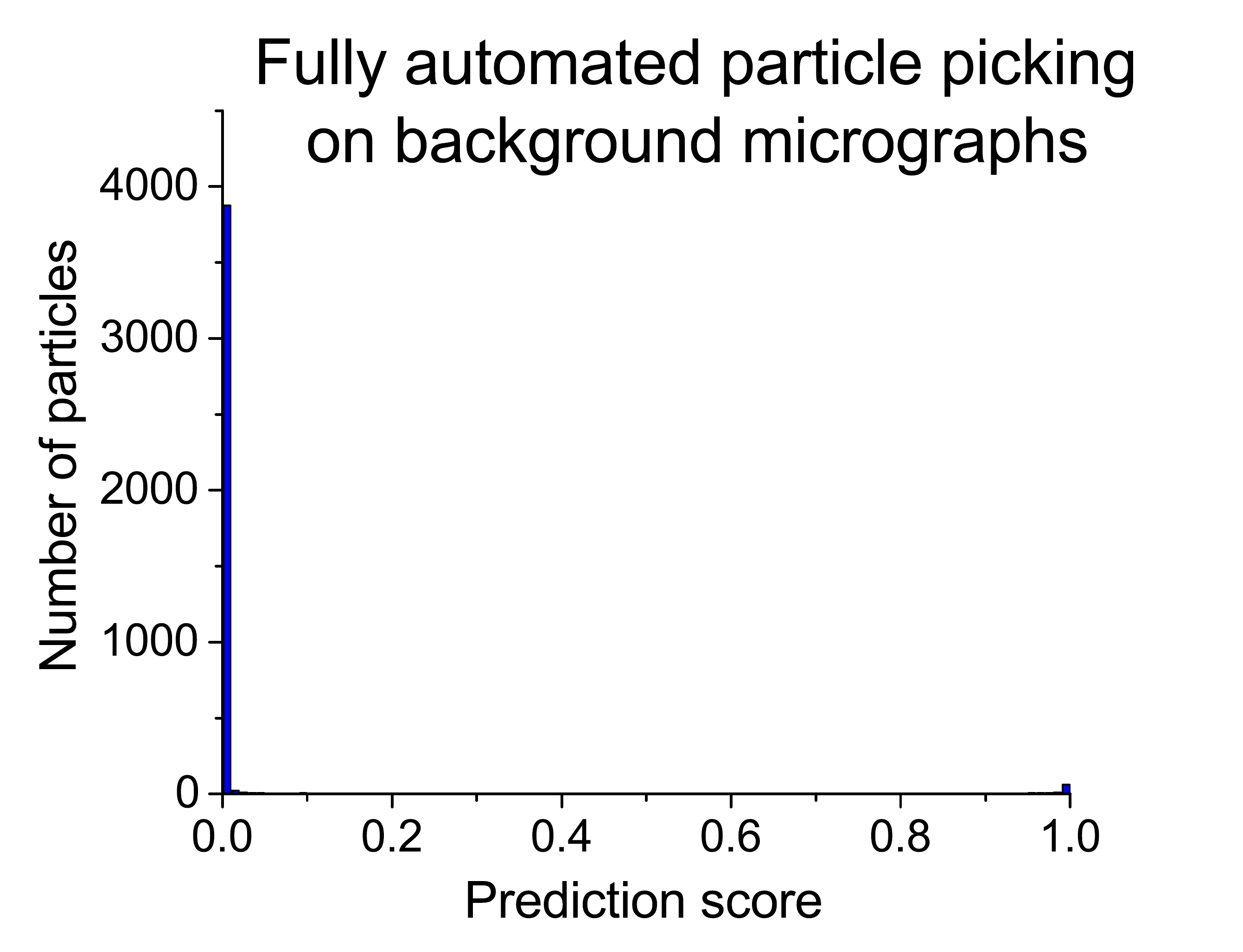}}
\subfigure[]{\includegraphics[width=5.2cm,height=4cm]{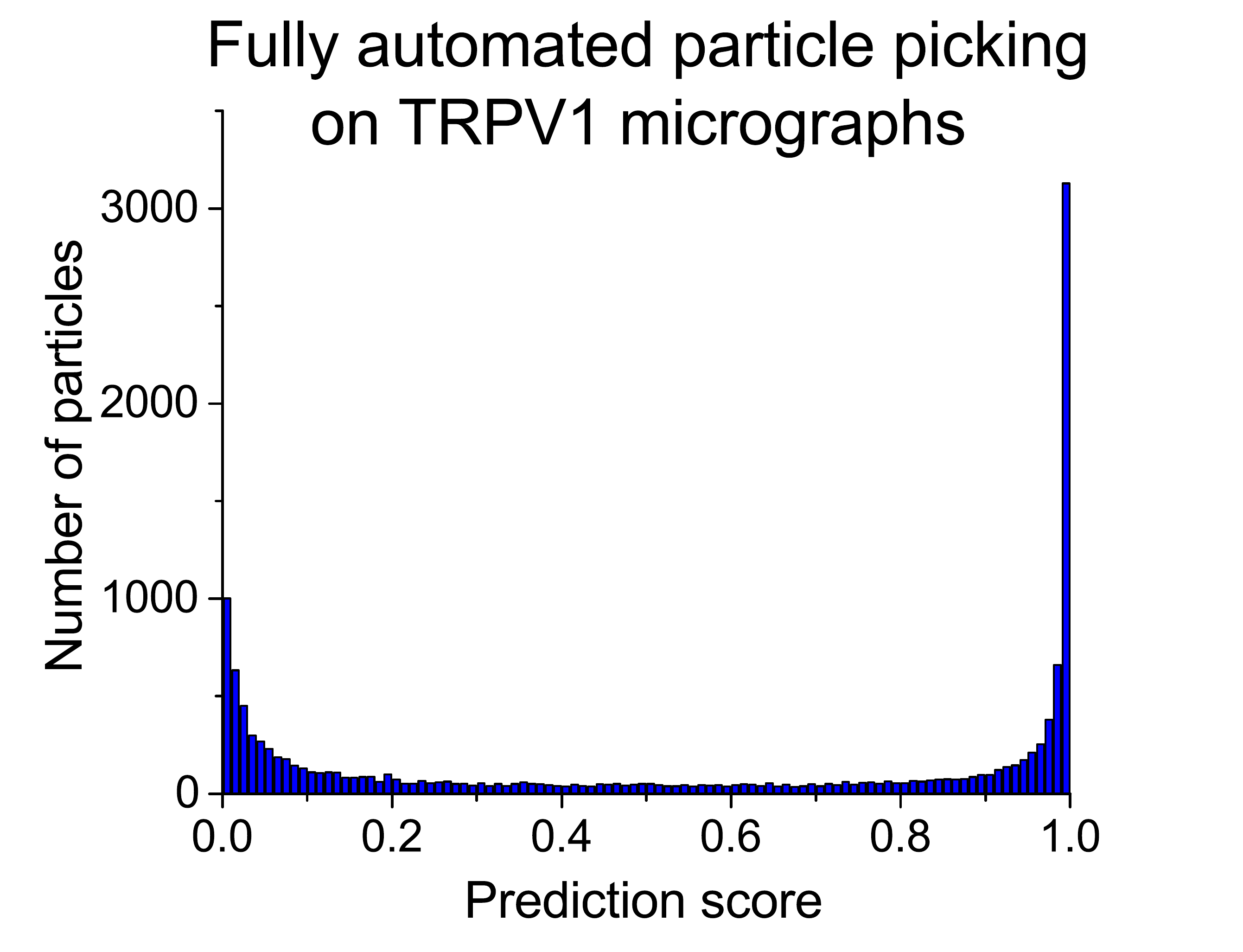}}
\caption{The results of fully automated particle picking from micrographs at different defocus levels and from random noise. (a), (b) and (c) The recall results of automated picking at different defocus levels for $\gamma$-secretase, spliceosome and TRPV1, respectively. The CNN classifier was trained by a mixture of 10000 particles of other four molecules that were different from the target one, each contributing to 2500 particles. The inlet panels in each subfigure show typical image examples at the minimum and maximum defocus levels, respectively. (d) and (e) The distributions of the prediction scores in our fully automated particle picking approach on pure background (i.e., without adding any sample) and TRPV1  micrographs, respectively.}\label{Discussion}
\end{figure}

The difficulty of picking particles from a micrograph is generally associated with the contrast level of the image, which is generally related to its defocus level. To examine the influence of the contrast levels on the automated particle selection results, we further examined the recall scores of our fully automated scheme at different defocus levels (Fig.~\ref{Discussion}). The tests on the micrographs of $\gamma$-secretase, spliceosome and TRPV1 with defocus ranging from 1 $\mu$m to 3.5 $\mu$m have shown that the performance of our fully automated particle picking method was relatively robust at different defocus levels (Fig.~\ref{Discussion} (a), (b) and (c)). This implies that our automated method can identify a sufficient number of correct particles at various conditions of data collection and thus may have a wide range of applications in single-particle cryo-EM data analysis.

Recently, it has been pointed out that the template-based particle picking methods might fall into a potentially dangerous pitfall~\cite{Henderson05112013,VanHeel2013,Subramaniam2013}, termed ``Einstein-from-noise''~\cite{Shatsky2009}, in which a photograph of Einstein was used as a template and pure noise images were aligned to this reference to reconstruct the Einstein image. Although our deep learning approach is essentially different from the template-based methods, it would be interesting to know whether our approach also suffers from this general problem, when using the cross-molecule particles as training data and picking particles from random noise. To investigate this issue, we tested our CNN model, which was trained by a mixture of 5000 $\gamma$-secretase and 5000 spliceosome particles, on 26 pure background micrographs which were measured from an empty cryo-EM instrument without putting any sample inside. In this additional test on background images with random noise, we found that it was difficult to pick particles with the normal threshold value as used in the previous tests of $\gamma$-secretase, spliceosome or TRPV1 (see Section 2). We then set the threshold to zero and investigated the distribution of the prediction scores of the selected particles. We found that for background with random noise, almost all the picked particles were associated with a low prediction score (less than 0.05) (Fig.~\ref{Discussion} (d)), which were greatly different from our previous results, e.g., the test of TRPV1 (Fig.~\ref{Discussion} (e)). This result indicates that very unlikely our automated picking process with the cross-molecule training scheme would fall into the previously-mentioned pitfall.

In addition to its own virtue in full automation, our deep learning framework with the cross-molecule training strategy has several additional advantages. First, unlike the subjective manual picking, the picking results of our deep learning classifier are objective and mainly determined by the in-house ``computational'' experts learned from training data. Second, as we apply a cross-molecule training method, in principle, we can incorporate as many available known particles from previously-solved complex structures as possible into training data, which will thus increase the robustness of the training process and further improve the automated picking results. Third, in our fully automated picking scheme under the deep learning framework, the cross-molecule training process can be finished before collecting micrograph data of the current target molecule. Such an offline training manner is much more efficient than the online training methods (i.e., the classifiers are trained after data collection) in most of existing classification based frameworks. Moreover, the deep learning framework can take full advantage of current available powerful computing platforms, e.g., graphics processing unit (GPU), to speed up the learning process. Thus, our fully automated picking approach can be generally much more efficient and more robust than previous semi-automated methods.

In summary, we have proposed a deep learning framework for automated particle picking. The powerful deep learning model can be applied to pick particles either in a semi-automated manner or through a fully automated fashion. In the fully automated particle picking strategy, we innovatively use the known particles of the previously solved structures to train the deep learning model, which can accurately describe the common and cross-molecule features of particles. Both our fully automated and semi-automated particle picking schemes have been tested on the published cryo-EM datasets of several molecules, including $\gamma$-secretase, spliceosome and TRPV1. The comparisons between the particles picked by either our fully automated or semi-automated method and those manually identified by human experts have demonstrated that our deep learning framework can achieve near human-level performance in particle picking. Thus, our fully automated or semi-automated particle picking approach can provide a reliable and practically useful tool to liberate structural biologists from the time-consuming and laborious task of manual particle picking in cryo-EM data analysis.

\vspace{-10pt}
\section{Materials and methods}

\subsection{Datasets and data preprocessing}

The data used in this paper were divided into two non-overlapping parts: a training dataset and test dataset. The training data consisted of 100, 400, 100, 84 and 400 micrographs of $\gamma$-secretase~\cite{GAMMA}, spliceosome~\cite{Spliceosome}, TRPV1~\cite{TRPV1}, $\beta$-galactosidase~\cite{Beta} and N-ethylmaleimide sensitive factor complex~\cite{SNARE}, respectively, while the test data contained a separate set of micrographs from  $\gamma$-secretase, spliceosome and TRPV1, each having 100 micrographs. All the results described in the paper are referred to those on the test datasets. The defocus value of each micrograph was calculated using CTFFIND4~\cite{CTF}. For each micrograph, we first used a Gaussian filter as a low pass filter to remove white noise with high frequency components. Then the binning strategy~\cite{Binning} was used to convert each original micrograph to an image ranging between 1,000 and 2,000 pixels. All the coordinates of reference particles were identified manually by human experts (see also Section 4.4) except for $\beta$-galactosidase, in which the coordinates of reference particles were obtained from EMPIAR (entry ID: EMPIAR-10017).

The $\gamma$-secretase and spliceosome datasets were obtained from Dr. Yigong Shi's lab at Tsinghua University, which were acquired by an FEI Titan Krios electron microscope operating at 300 kV with a magnification of 22,500x. These micrographs were first taken by a Gatan K2 Summit using a super-resolution mode at 0.66 \AA/pixel, and then binned to a final pixel size of 1.32 \AA. The TRPV1 dataset was downloaded from EMPIAR (entry ID: EMPIAR-10005), in which the micrographs were recorded by a Gatan K2 Summit equipped with a FEI POLARA 300 operating at 300 kV in a super-resolution counting mode, with a final calibrated super resolution pixel size of 1.2156 \AA. The $\beta$-galactosidase dataset was downloaded from EMPIAR (entry ID: EMPIAR-10017), which was recorded by a FEI Falcon-\uppercase\expandafter{\romannumeral2} camera. More detailed information of the $\beta$-galactosidase data collection can be found in~\cite{Beta}. The NSF (N-ethylmaleimide Sensitive Factor) fusion complex (20S particle) dataset was kindly offered by Dr. Minglei Zhao from Dr. Axel T. Brunger's Lab, and more details about its data collection can be found in the original paper~\cite{SNARE}.

\subsection{The convolutional neural network model}

Our deep learning framework employs a convolutional neural network (CNN) as the classifier to discriminate correct particles (i.e., positive samples) from random noise images (i.e., negative samples). Here, we briefly describe the concept of a CNN. The reader is referred to~\cite{CNN1990,Lecun98gradient-basedlearning,Reviewdeep,ImageNet,Facerecognition} for more details of this deep learning model.

A convolutional neural network is a specific type of deep learning~\cite{ImageNet,Lecun98gradient-basedlearning, VUCN} with a multi-layer structure, which consists of an input layer, an output layer and several hidden layers (between input and output layers). Each layer is comprised of a number of units, also called artificial neurons. In a CNN, the output layer is a softmax layer~\cite{PRML} in which the number of artificial neurons is equivalent to that of the classification classes, and the value of each unit represents the chance of being the corresponding class. In a softmax layer, the output of a CNN is ensured to be a probability value between 0 and 1. More specifically, a softmax function is defined as $$P(Y_{i}) = \frac{e^{W_{i}X+b_{i}}}{\sum_{j}e^{W_{j}x+b_{j}}}$$ where $Y_{i}$ represents the output, $i$ and $j$ stand for the indexes of the artificial neurons in the last layer (which also represent the indexes of the output classes), $\bm{X}$ represents the input to the softmax layer, $W_{i}$ represents the corresponding weight parameter, and $b_{i}$ stands for the bias of the $i^{th}$ unit in the output layer.

The hidden layers mainly include convolutional, pooling and fully-connected layers. The convolutional and fully-connected layers are also called the learning layers, since their weights are optimized during the training procedure. The dimension of the weight parameters in a convolutional layer is $N \times M \times K \times K$, where $N$ stands for the number of kernels in the convolutional layer, $M$ stands for the number of square matrices in the kernel (e.g., a color image has three square matrices), and $K$ represents the dimension of a square matrix which is typically less than 10. The convolutional layer takes the convolution operation between the input $\bm{X}$ and the weights $\bm{W}$ and then feeds the output into a nonlinear function. In particular, the convolutional operation is defined as
$$convolution(X)_{ijn} = F_{nonlinear}\left(\sum_{m=0}^{M-1}\sum_{k=0}^{K-1}\sum_{l=0}^{K-1}W_{mkl}^{n}X_{m,k+i-1,l+j-1}\right)$$
where $i$ and $j$ denote the location indexes of the output, $n$ is the index of the output kernel, $m$ is the index of the input matrix, $k$ and $l$ stand for the location indexes of the kernel, and $F_{nonlinear}$ stands for a nonlinear function, e.g., $tanh(x)$, $sigmoid(x)=\frac{1}{1+e^{-x}}$ and the rectified linear function (ReLU) $ReLU(x) = \max(0,x)$. The ReLU function has been popularly used in a CNN architecture, as it can significantly speed up the training process~\cite{ImageNet, ReLU}.

The pooling layer first divides the input image into non-overlapping windows, and then takes a max or mean operation over each window. The pooling operation can reduce the number of parameters and thus the complexity of the learning model. In addition, it can effectively relieve the over-fitting problem~\cite{Jarrett2009,Scherer2010}.

The architecture of our CNN model consists of four convolutional layers, each of which uses an ReLU function and is followed by a max pooling layer. Then these convolutional and max pooling layers are stacked with two fully-connected layers. The input data of our CNN framework is a patch image within sliding window (see Section 2.1), and the results of the last fully-connected layer are fed into a softmax layer with two units, which represent the positive and negative labels, respectively. The ``dropout" technique~\cite{Dropout}, in which the unit states in a layer are set to zero with probability $p$ (a typical value is 0.5), is applied in the first fully-connected layer to address the overfitting problem~\cite{ImageNet, Dropout}.

\subsection{Training}

The stochastic gradient descent algorithm~\cite{Bottou} is used to train our deep learning model. The training data to our deep learning classifier include a set of positive samples and an equivalent number of negative samples, which are selected from those regions at least 0.6 times the size of sliding window away from positive samples. The positive samples are either from the known particles of other molecules whose structures have been previously determined via cryo-EM (i.e., in a fully automated manner) or the manually picked particles of the target molecule (i.e., in a semi-automated manner). As particles may have different sizes, we normalize both positive and negative samples into images of 64 pixels $\times$ 64 pixels. This normalization scheme is particularly useful when we use cross-molecule data to train our deep learning model, as the particles of distinct molecular complexes generally have different sizes. During the training process, we also use a separate validation dataset, which is independent from training data and test data, to determine some super-parameters of our deep learning model, e.g., the number of iterations to reach the convergence. The size of such a validation dataset is chosen to be around 1/9 of the training dataset.

In our fully automated picking scheme, we first use a mixture of 10,000 known particles of other molecules as training data that were different from the target complex, each contributing to a roughly equivalent number of training examples. Then in the iteration step (Fig.~\ref{Methods}), the top 10,000 particles predicted by the previously trained deep learning classifier are used as new training data to further refine the deep learning model. For all the test results shown in Section 2.2, we only ran one iteration of the training process. We did not observe significant change when conducting more training iterations.

In the cross-molecule training strategy, we also tested different combinations of the known particles from other molecules as training data. In particular, we looked into the test results on the deep learning classifier trained by the known particles of one, two, three and four types of molecules. Our paper only shows the typical results of each case (Fig.~\ref{AutoPR}). More details about different combinations of training data are provided in Table.~\ref{Auto}. The tests of other combinations show similar results for individual cases.

\subsection{Performance evaluation}

We evaluated the performance of our particle picking approach mainly by comparing the automated picking results to the corresponding reference particles picked manually by human experts. These reference particles were first picked by one cryo-EM expert and then further verified by two additional experts. The coordinates of all manually-picked particles were further aligned using FREALIGN~\cite{
FREALIGN}. An automatically-picked particle was said to agree with a manually picked particle, if the distance between their centers was less than a threshold, which was set to be 0.2 times the size of sliding window. We mainly used recall and precision to evaluate the accuracy of the particles selected by our automated approach. Let $TP$ denote the number of particles that were picked by our deep learning approach and also agreed with reference particles picked by human experts. Let $FP$ denote the number of particles that were picked by our approach but did not agree with any reference particle. Let $FN$ denote the number of the reference particles that were picked by human experts but not by our automated approach. Then recall and precision are defined as
$recall = TP/(FN + TP)$, and $precision = TP/(FP + TP)$, respectively.

In addition to recall and precision, we also measured the deviations of the centers of the particles picked by our automated approach from those of the reference particles selected by human experts. In particular, we assessed their average distances, normalized by the size of sliding window.

We also compared the 2D clustering and class averaging results of the particles picked by our approach to those from the manually picked images in the reference set. The 2D clustering and class averaging operations were performed using the corresponding commands in RELION 1.3. Such a 2D analysis process was conducted in 2-3 iterations. In each iteration, those obvious image artifacts, invalid particles or empty objects from the 2D averaging results were removed from the candidate list. The same data analysis procedure was done for both manually and automatically picked particles.

\subsection{Parameter setting}

In this section, we explain the parameter setting of our deep learning model for automated particle picking. In the data preprocessing stage, the standard deviation of a gaussian filter was set to 0.1. In the model training stage, the learning rate was set to 0.01. In the scoring step of the particle picking stage, the size of sliding window was set to be 180 pixels for TRPV1 and $\gamma$-secretase, and 320 pixels for spliceosome, and the step size was set to be 4 pixels. In the cleaning step, two neighboring pixels with prediction scores above 0.5 were connected. In addition, we removed those connected domains in which the size was four times larger than the average size. In the filtering step, the size of peak window was set to be 0.8 times the size of sliding window, in other words, the minimum distance between centers of two potential particles was set to be 0.4 times the size of sliding window. When removing the bad particles, the extreme pixels are defined as those with three standard deviations away from the mean pixel value. In the iteration step, we chose 10,000 samples from the top sorted particles picked by the previously-trained classifier to further refine the model.

\subsection{Implementation}
DeepPicker was implemented based on Torch7 (\url{http://torch.ch}), an open source deep learning library and the Lua programming language. It can be run either with or without graphic processing units (GPUs). In our tests, a NVIDIA Quadro K4000 GPU was used for training the model and picking the particles from micrographs. The time required for training the model with 10,000 samples was less than an hour, and picking the particles from a micrograph of size 3710 pixels $\times$ 3838 pixels took about 1.5 minutes. DeepPicker will be available and distributed as an open-source program.

\vspace{-10pt}
\section{Acknowledgements}

We thank Prof. Yigong Shi, Prof. Nieng Yan, Dr. Minglei Zhao and Mr. Jianping Wu for generously sharing their cryo-EM data for testing our particle picking approach. We are grateful to Prof. Hongwei Wang, Prof. Jian Peng and Mr. Jinglin Yang for their insightful discussions about our work. We are also grateful to Ms. Rudan Chen for her help on our draft preparation. Part of the test data are from EMPIAR (\url{https://www.ebi.ac.uk/pdbe/emdb/empiar/}).

\section{Additional information}

\subsection{Funding}
This work was supported by the National Basic Research Program of China (Grant 2011CBA00300, 2011CBA00301 to JZ), the National Natural Science Foundation of China (Grant 61033001, 613611\\36003 and 61472205 to JZ, 31570730 to XL, and 61571202 to TX), China Youth 1000-Talent Program by the State Council of China (to JZ, XL and TX), Beijing Advanced Innovation Center for Structural Biology (to JZ and XL), and Tsinghua-Peking Joint Center for Life Sciences (to XL). We acknowledge the China National Center for Protein Sciences Beijing for providing the facility support and the support of NVIDIA Corporation with the donation of the Titan X GPU used for this research.

\subsection{Author contributions}

XL and JZ initiated the project. JZ and XL designed the functions of the program. JZ, TX, FW and HG designed the deep learning algorithm. FW and HG implemented the program. FW, HG, GL and ML performed the testing experiment and analysis. CY acquired part of the data and provided corresponding reference images. JZ, HG and FW drafted the manuscript. JZ, XL and TX revised the manuscript.

\newpage
{

}

\clearpage
\begin{table}[hbp]
\begin{tabular}{|c|c|}
\hline
\multicolumn{2}{|c|}{\leftline{For $\gamma$-secretase:}}\\
\hline
1 type & TRPV1 \\
\hline
2 types & TRPV1 and spliceosome \\
\hline
3 types & TRPV1, spliceosome and NSF fusion complex\\
\hline
4 types & TRPV1, spliceosome, NSF fusion complex and $\beta$-galactosidase\\
\hline
\multicolumn{2}{|c|}{\leftline{For spliceosome:}}\\
\hline
1 type & $\gamma$-secretase \\
\hline
2 types & $\gamma$-secretase and TRPV1 \\
\hline
3 types & $\gamma$-secretase, TRPV1 and NSF fusion complex\\
\hline
4 types & $\gamma$-secretase, TRPV1, NSF fusion complex and $\beta$-galactosidase\\
\hline
\multicolumn{2}{|c|}{\leftline{For TRPV1:}}\\
\hline
1 type & spliceosome \\
\hline
2 types & spliceosome and $\gamma$-secretase \\
\hline
3 types & spliceosome, $\gamma$-secretase and NSF fusion complex\\
\hline
4 types & spliceosome, $\gamma$-secretase, NSF fusion complex and $\beta$-galactosidase\\
\hline
\end{tabular}
\caption{Different combinations of training data for the test results shown in Fig.~\ref{AutoPR}.}\label{Auto}
\end{table}

\end{document}